\documentclass[twocolumn]{aastex631}
\usepackage{gensymb}
\usepackage[T1]{fontenc} 
\usepackage[utf8]{inputenc}

\shorttitle{GCM interpretation of WASP-121 $\rm{b}$'s NIRISS Phase Curve}

\newcommand{\unit}[1]{\ensuremath{\, \mathrm{#1}}}

\newcommand{\UMICH}{Department of Astronomy and Astrophysics, University of Michigan, Ann Arbor, MI, 48109, USA}

\newcommand{\MCGILL}{Department of Earth and Planetary Sciences, McGill University, 3450 rue University, Montr\'eal, QC H3A OE8, Canada}

\newcommand{\TROMC}{Trottier Space Institute at McGill, 3550 rue University, Montr\'eal, QC H3A 2A7, Canada}

\newcommand{\MONTREAL}{Institut Trottier de recherche sur les exoplan\`etes, Département de Physique, Universit\'e de Montr\'eal, Montr\'eal, Québec, Canada}

\begin{document}

\title{The Days Drag On on WASP-121 b: Interpreting its NIRISS Spectroscopic Phase Curve with General Circulation Models}

\correspondingauthor{Robert C. Frazier}
\email{robertcf@umich.edu}

\author[0000-0001-6569-3731]{Robert C. Frazier}
\affil{\UMICH}

\author[0000-0003-3963-9672]{Emily Rauscher}
\affil{\UMICH}

\author[0000-0001-9987-467X]{Jared Splinter}
\affil{\TROMC}
\affil{\MCGILL}

\author[0000-0002-2984-3250]{Thomas D. Kennedy}
\affil{\UMICH}

\author[0000-0003-2278-6932]{Xianyu Tan}
\affil{Tsung-Dao Lee Institute, Shanghai Jiao Tong University, 520 Shengrong Road, Shanghai, People’s Republic of China}
\affil{School of Physics and Astronomy, Shanghai Jiao Tong University, 800 Dongchuan Road, Shanghai, People’s Republic of China}

\author[0000-0001-9521-6258]{Vivien Parmentier}
\affil{Université Côte d’Azur, Observatoire de la Côte d’Azur, CNRS, Laboratoire Lagrange, Nice, France}

\author[0000-0003-0217-3880]{Isaac Malsky}
\affil{Jet Propulsion Laboratory, California Institute of Technology, Pasadena, CA 91109, USA}

\author[0000-0002-2195-735X]{Louis-Philippe Coulombe}
\affil{Plan\'etarium de Montr\'eal, Espace pour la Vie, 4801 av. Pierre-de Coubertin, Montr\'eal, Canada}
\affil{\MONTREAL}

\author[0000-0002-1199-9759]{Romain Allart}
\affiliation{\MONTREAL}

\author[0000-0001-6129-5699]{Nicolas B. Cowan}
\affil{\TROMC}
\affil{\MCGILL}
\affil{Department of Physics, McGill University, 3600 rue University, Montréal, QC H3A 2T8, Canada}

\author[0000-0002-6780-4252]{David Lafreni\`{e}re}
\affil{\MONTREAL}

\author[0000-0003-4816-3469]{Ryan MacDonald}
\affil{School of Physics \& Astronomy, University of St Andrews, North Haugh, St Andrews, KY16 9SS, UK}

\author[0000-0002-8573-805X]{Stefan Pelletier}
\affiliation{Observatoire astronomique de l'Université de Genève, 51 chemin Pegasi 1290 Versoix, Switzerland}

\author[0000-0003-4987-6591]{Lisa Dang}
\affil{\MONTREAL}
\affiliation{Department of Physics and Astronomy, University of Waterloo, 200 University W, Waterloo, Ontario, Canada, N2L 3G1}

\author[0000-0001-5485-4675]{Ren\'{e} Doyon}
\affil{\MONTREAL}

\author[0000-0002-6773-459X]{Doug Johnstone}
\affiliation{\textnormal{NRC Herzberg Astronomy and Astrophysics, 5071 West Saanich Rd, Victoria, BC, V9E 2E7, Canada }}
\affiliation{\textnormal{Department of Physics and Astronomy, University of Victoria, Victoria, BC, V8P 5C2, Canada}}

\author[0000-0002-0436-1802]{Lisa Kaltenegger}
\affiliation{Department of Astronomy and Carl Sagan Institute, Cornell University, Ithaca, NY 14853, USA}

\author[0000-0003-1227-3084]{Michael R. Meyer}
\affil{\UMICH}

\author[0000-0002-2875-917X]{Caroline Piaulet-Ghorayeb}
\altaffiliation{E. Margaret Burbridge Postdoctoral Fellow}
\affiliation{Department of Astronomy \& Astrophysics, University of Chicago, 5640 South Ellis Avenue, Chicago, IL 60637, USA}

\author[0000-0002-3328-1203]{Michael Radica} 
\altaffiliation{NSERC Postdoctoral Fellow}
\affiliation{Department of Astronomy \& Astrophysics, University of Chicago, 5640 South Ellis Avenue, Chicago, IL 60637, USA}

\author[0000-0001-7836-1787]{Jake D. Turner}
\affiliation{Department of Astronomy and Carl Sagan Institute, Cornell University, Ithaca, NY 14853, USA}

\begin{abstract}

Ultra-hot Jupiters present extreme atmospheric phenomena not found in the Solar System. These planets' daysides experience strong temperature inversions, molecular species (including H$_2$) dissociate, and magnetism disrupts their atmospheric circulation. On their nightsides H$_2$ can recombine and clouds may form. Spectroscopic phase curves let us measure these spatially inhomogeneous conditions, which can then be interpreted with three-dimensional (3-D) models. In this work we compare the JWST/NIRISS spectroscopic phase curve of the ultra-hot Jupiter WASP-121 b to state-of-the-art 3-D models with varying modeling assumptions, including the aforementioned physical phenomena. 
\added{We demonstrate the importance of accurately accounting for the planet's radius in comparison between data and models, as it changes the implied overall planetary emission. We find that the 3-D models predict planet emission $\sim$12\% higher than observed, contributing to a continued tension between measured and predicted hot Jupiter albedos.} We identify multiple pieces of evidence that confirm a strong source of drag operating in this planet's atmosphere. In addition, the nightside emission spectrum is devoid of strong absorption features, which may be best explained by nightside clouds. One feature of the dataset that is not matched by the 3-D models is a trend of increasing eastward phase offset with decreasing wavelength, for wavelengths shorter than $\sim$1.4 $\mu$m. This result is not consistent with reflection from dayside clouds, nor can it be explained by removing atmospheric opacity sources. Our analysis highlights the complexities in generating 3-D models and interpreting observations of ultra-hot Jupiters in the JWST era.
\end{abstract}


\section{Introduction} \label{sec:intro}

Among the plethora of exoplanets discovered so far, ultra-hot Jupiters \citep[UHJs, $\mathrm{T_{eq} >}$ 2200 K;][]{parmentier_thermal_2018} exhibit some of the most extreme atmospheric physics. Hot Jupiters are expected to have strong day-night temperature gradients ($\Delta T \gtrsim 1000 \unit{K}$) between their perpetual day- and night-sides because they should be tidally locked into synchronous rotation states but, even beyond this, the dayside atmospheres of UHJs should be hot enough for many molecules to thermally dissociate \citep{parmentier_thermal_2018, Arcangeli2018}. As a consequence, species, such as H$_2$O, are removed as sources of opacity, weakening the strength of these molecular emission features \citep{parmentier_thermal_2018}. Additionally, the introduction of H$^-$ opacity can further change the radiative balance and dampen spectral features, especially at wavelengths shorter than 1.6$\unit{\mu m}$. The dissociation of H$_2$ on the dayside of UHJs can act as a significant heat sink and its recombination on the nightside an important heat source, improving the heat-recirculation in the planet's atmosphere \citep{bell_increased_2018}. This impacts the driving of the winds that transport gas from the day-side to night-side and shape the overall global temperature structure \citep{tan_modelling_2024}. The three-dimensional properties of UHJ atmospheres can be further complicated by magnetohydrodynamics influencing circulation on the dayside, where there is a significant level of thermal ionization \citep{perna_magnetic_2010,rogers_magnetic_2014,rogers_constraints_2017,coulombe_broadband_2023, beltz_comparative_2024}, the presence of mineral or metallic clouds forming on the cooler nightside \citep{roman_modeled_2019,roman_clouds_2021,komacek_patchy_2022, bell_nightside_2024, coulombe_highly_2025}, or both of these effects together \citep{kennedy_radiatively_2025}. All these phenomena lead to an incredibly complex and intrinsically three-dimensional picture of an UHJ’s atmosphere and make them a prime subject for observations.

Among the most well studied UHJs is WASP-121 b \citep[$T_{\mathrm{eq}} = 2409$ K, R$_{\rm{p}}$ = 1.742 $R_J$, P = 1.27 days, M$_{\rm{p}}$ = 1.17 $M_J$;][]{Sing2024}. Since its discovery \citep{delrez_wasp-121_2016} it has been the target of a wide range of observations which have looked to understand its complex atmosphere. A Hubble Space Telescope (HST) Wide Field Camera 3 (WFC3) transmission spectrum from \cite{evans_detection_2016} indicated the presence of water in the planet's atmosphere and \cite{evans_ultrahot_2017} found the presence of water emission lines in the dayside spectrum of the planet with HST/WFC3, identifying a temperature inversion in the upper atmosphere. These thermal inversions are thought to be driven by strong optical absorbers such as atomic metal species, TiO, and VO \citep{fortney_unified_2008,parmentier_thermal_2018, lothringer_extremely_2018}, but ground based high resolution spectroscopy has proven inconclusive in finding TiO and VO in WASP-121 b's atmosphere. A more global view of this planet has come from phase-curve observations with TESS \citep{bourrier_optical_2020,daylan_tess_2021}, HST/WFC3 \citep{mikal-evans_diurnal_2022, changeat_2024}, Spitzer/IRAC \citep{morello_spitzer_2023,dang_comprehensive_2025, davenport_analysis_2025}, JWST/NIRSpec \citep{mikal-evans_jwst_2023, Sing2024}, and most recently, JWST NIRISS \citep{Splinter2025}.
These phase curve measurements have generally found little-to-no offset between the brightest region of the planet and the sub-stellar point, although the measured offset values differ between observations, even within the same wavelength range or for different analyses of the same dataset.
A small phase offset is consistent with drag in the atmosphere impeding heat recirculation. \cite{mikal-evans_diurnal_2022} presented spectroscopic phase curves of the planet from HST/WFC3 and found that the nightside of the planet lacks a thermal inversion and is cool enough that we should expect cloud species to condense \citep[a finding supported by][]{mikal-evans_jwst_2023}, but was unable to unambiguously identify the presence of clouds.

The recent JWST/NIRISS (Near Infrared Imager and Slitless Spectrograph) observations of WASP-121~b provide the planet’s spectroscopic phase curve using the SOSS (Single Object Slitless Spectroscopy) mode \citep[0.6-2.85$\unit{\mu m}$, R $\sim$700 at 1.4$\unit{\mu m}$;][]{albert_near_2023, doyon_niriss_2023}. With a greater spectral resolution and wider wavelength coverage than HST/WFC3, we can use these data to better understand the three-dimensional (3-D) atmospheric structure of this planet. The NIRISS spectroscopic phase curves were presented, along with an initial analysis in \cite{Splinter2025}. Here we present a detailed comparison between these spectroscopic phase curves and a set of simulations produced from state-of-the-art 3-D General Circulation Models (GCMs) to gain insight into the extreme physics operating in this planet's atmosphere. 

In Section \ref{sec:meth} we provide details of the JWST NIRISS/SOSS spectroscopic phase curve of this planet and describe the GCMs, and their outputs we compare against the data, including how we calculated their simulated phase curves. In Section \ref{sec:results} we compare the simulated spectroscopic phase curves to the NIRISS observations to constrain the physical state of WASP-121 b's atmosphere. In Section \ref{sec:disc} we discuss sources of uncertainty in our modeling and identify what remains to be understood about this planet. The results of this work are summarized in Section \ref{sec:conc}.

\section{Methods} \label{sec:meth}

The phase curves presented in this work were reduced and analyzed by \cite{Splinter2025}. The data were taken with JWST NIRISS (Near Infrared Imagining Slitless Spectrograph) in SOSS (Single Object Slitless Spectroscopy) mode, which covers a wavelength range of 0.6-2.85$\unit{\mu m}$ \citep{albert_near_2023}. The observation captured the full phase curve of WASP-121 b, including a transit and two secondary eclipses, which are analyzed separately in the companion papers: \cite{Pelletier2025} and MacDonald et al., in prep. The observation lasted for 36.90 hours with 6 groups per integration for a total of 3452 integrations. This dataset is part of the NIRISS Exploration of the Atmospheric Diversity of Transiting Exoplanets (NEAT) Guaranteed Time Observation Program (GTO 1201: PI D. Lafreniere). 

\cite{Splinter2025} presents two independent reductions of the data, one performed with \texttt{exoTEDRF} \citep{feinstein_early_2023, radica_awesome_2023, radica_exotedrf_2024} and one with \texttt{NAMELESS} \citep{coulombe_broadband_2023, coulombe_highly_2025}. We primarily use the \texttt{exoTEDRF} version as it is presented as the fiducial reduction in \cite{Splinter2025}. The data are split between Order 1 (0.85-2.85$\unit{\mu m}$) and Order 2 (0.6-1.4$\unit{\mu m}$), but the reduction only used the 0.6-0.85$\unit{\mu m}$ range of Order 2. The \texttt{exoTEDRF} reduction fit the phase curve using a model that accounted for the tidal deformation that WASP-121 b experiences throughout its orbit \citep[although this effect is expected to be minimal,][]{delrez_wasp-121_2016}, stellar variability on Order 2 of NIRISS/SOSS, and using a prior that \added{penalizes negative} planetary flux. The spectral $F_p/F_s$ light curves produced by the reduction and used in this work are binned into 5 pixel bins and then normalized to the first eclipse. The final data product has a total of 522 spectroscopic phase curves, 113 from Order 2 and 409 from Order 1. For a more in-depth discussion of this as well as the overall analysis of the phase curve and energy budget of WASP-121 b see \cite{Splinter2025}.

\subsection{3-D Planet Models}

We compare the NIRISS/SOSS phase curve of WASP-121~b to a set of 3-D models, using two separate GCM codes (the RM-GCM and the SPARC/MITgcm) and varying assumptions for the physical processes at work in the planet's atmosphere.

\subsubsection{RM-GCM General Circulation Model \label{sec:rmgcm}}

Originating from an Earth GCM modeling system \citep[the University of Reading’s Intermediate General Circulation Model;][]{hoskins_multilayer_1975}, the RM-GCM was adapted for simulations of hot Jupiter atmospheres \citep{rauscher_three-dimensional_2010}. The dynamical core of the RM-GCM solves the primitive equations of meteorology in pseudo-spectral space and it is coupled to a two-stream radiative transfer routine \citep{toon_rapid_1989}. This work uses version 5.0 of the open-source RM-GCM \citep{rauscher_rm-gcm_2023} with picket fence radiative transfer, described in detail in \cite{malsky_direct_2024} and \cite{kennedy_radiatively_2025}. The picket-fence coefficients used can be found in \cite{parmentier_non-grey_2015}, which assume equilibrium chemistry and include contributions from TiO and VO.
The opacity sources used to calculate them can be found in \cite{freedman_gaseous_2014}.

The RM-GCM uses a kinematic magnetohydrodynamics approach to capture the interaction between the partially thermally ionized atmosphere and a global magnetic field (assumed to be generated in the interior), as detailed in \cite{rauscher_three-dimensional_2013} and \cite{beltz_exploring_2022}. Under the assumptions used, the east-west winds experience Lorentz drag from the interaction of embedded ions with the planetary magnetic field. Kinetic energy removed via this drag is added back locally as ohmic heating \citep{rauscher_three-dimensional_2013}. The strength and timescale of this interaction is set by both the strength of the global magnetic field (which is set to $B=3$ G in this work) as well as the local ionization state in the atmosphere. The ionization state is actively calculated with the Saha equation as the model runs \citep{menou_magnetic_2012}. Simplifications are made to this model such as: 1) a limit to the lowest magnetic drag timescale used, in order to maintain computational stability, \added{2) having the Lorentz forces act only in the zonal direction, as winds in non-dragged UHJ models are primarily zonal \citep[although see recent work by ][ that shows how relaxing these assumptions modifies the predicted circulation in quantitatively significant ways, but without changing the overall global pattern at a qualitative level]{Christie2025}}, and 3) neglecting more complex magnetohydrodynamics effects, such as the fact that the atmosphere can induce its own magnetic field component \citep{rogers_constraints_2017}. 

Cloud modeling in the RM-GCM is done with radiatively active clouds which form and dissipate throughout the simulation depending on if the local temperature and pressure are below a cloud's condensation curve. The clouds actively influence the radiative transfer within the model; scattering and absorption values are interpolated from tables that were pre-calculated with Mie theory. This cloud prescription is described in more detail in \cite{roman_modeled_2019}, \cite{roman_clouds_2021}, and \cite{malsky_direct_2024}. In this work we include eight possible cloud species: $\mathrm{KCl, Cr, SiO_2, Mg_2SiO_4, VO, Ca_2SiO_4, CaTiO_3,}$ and $\mathrm{Al_2O_3}$ and assume a ``compact'' configuration for the clouds where they are limited by the number of layers (10) they can extend above the cloud base ($\sim3$ scale heights in the atmosphere) \added{Large particles (like those assumed here) are expected to settle efficiently under the influence of gravity, particularly when the cloud base is in the upper atmosphere \citep{Lines2018}. Together, these justify our assumption of a compact cloud deck, but testing in our model has shown that the impact of varying cloud thickness on phase curves is minimal in the ultra-hot Jupiter regime \citep{kennedy_radiatively_2025}}.

The models produced from the RM-GCM that are presented in this paper were generated with a T31 horizontal resolution (corresponding to 96 longitudes and 48 latitudes), 50 vertical layers logarithmically spanning from $100-10^{-4}$ bar, and $9600$ time-steps per day.

\subsubsection{SPARC/MITgcm General Circulation Model} \label{sec:sparc}

The SPARC/MITgcm, as first described in \cite{showman_atmospheric_2009}, is adapted for modeling hot Jupiters from the MITgcm, a modeling system for Earth's atmosphere and oceans \citep{adcroft_implementation_2004}. The SPARC/MITgcm solves the primitive equations of meteorology on a cubed-sphere grid and uses a correlated-k two-stream radiative transfer routine \citep{marley_thermal_1999}.

The \cite{tan_modelling_2024} version of SPARC/MITgcm employs a hydrogen dissociation and recombination method implemented and discussed in detail in \cite{tan_atmospheric_2019} \added{and updated in \cite{komacek_patchy_2022} to incorporate inert He into the mass budget which can mute the impact of the effect}. H$_2$ dissociation is predicted \citep{bell_increased_2018} and observed \citep{mansfield_evidence_2020} to have a significant impact on the atmospheres of UHJs. These planets' daysides are hot enough that H$_2$ is expected to dissociate, a very endothermic process \citep[q = $2.14 \times 10^8 \unit{J \ kg}^{-1}$,][]{dean1999} that cools down the dayside and, as the winds on the planet advect atomic hydrogen to the nightside, it recombines and heats up the nightside \citep{tan_atmospheric_2019}. The code uses active tracers in the model to track the evolution of the atomic hydrogen mixing ratio which changes around the planet. As this mixing ratio changes, heat is released or absorbed due to, respectively, the recombination or dissociation of hydrogen. 

The SPARC/MITgcm also includes an optional linear drag on the horizontal momentum of the model, as implemented in \cite{komacek_atmospheric_2016}. This drag has a prescribed timescale that looks to, with computational simplicity and efficiency, replicate the impacts of drag effects like Lorentz drag and drag from turbulence. This is in addition to a basal drag\added{, motivated from a drag scheme first implemented in \citep{liu_atmospheric_2013},} that works to capture momentum exchange with the interior and aids in model stability and convergence. It has a timescale linearly increasing from $10^5$ s at the model's bottom pressure level to infinity (so no drag) at 20 bars \citep{tan_modelling_2024}.

The models produced from the SPARC/MITgcm that are presented in this paper were generated with a C32 resolution (corresponding to 128 longitudes and 64 latitudes) and 53 vertical layers spanning $200-10^{-6}$ bar.

\subsubsection{GCM Suite of Models \label{sec:prescriptions}}

In this work we use \added{nine} different models for the planet WASP-121~b, four run using the RM-GCM and five using the SPARC/MITgcm. The system parameters used to generate these models are listed in Table \ref{tab:stellarparam} and a summary of the full set of models is given in Table \ref{tab:gcmpres}. Note that the SPARC/MITgcm models use two separate versions of that code; one model is from \cite{parmentier_thermal_2018}, \added{three} are from \cite{wardenier_phase-resolving_2024}, using the code described in \cite{tan_modelling_2024}, \added{and an additional model produced for this work using this same code but no H$_2$ dissociation nor drag}. Importantly, the opacity sources used in the two versions of SPARC/MITgcm are not identical. Specifically, the version from \cite{wardenier_phase-resolving_2024} includes additional gaseous opacity sources in the radiative transfer that the version from \cite{parmentier_thermal_2018} does not; \added{the most significant of these is the inclusion of Fe which causes a slight decrease of the dayside photosphere temperature of the \cite{wardenier_phase-resolving_2024} models}. See Section \ref{sec:postprocess} for further details on the opacity sources used in the models. The models of WASP-121~b using the RM-GCM were presented in \cite{davenport_analysis_2025}; similar models of this planet using the RM-GCM were presented by \cite{beltz_comparative_2024} and a more detailed discussion of how the RM-GCM includes both clouds and magnetic drag can be found in  \cite{kennedy_radiatively_2025}. Here we summarize the global patterns resulting from these models, before comparing them to the JWST phase curve.

\begin{deluxetable}{lc}
\tablecaption{System Parameters \label{tab:stellarparam}}
\tabletypesize{\scriptsize}
\tablehead{\colhead{Parameter}       & \colhead{Value}   }
\startdata
\multicolumn{2}{c}{\hspace{-0.3cm} WASP-121}           \\
\hline
$R_*$                   & $1.458 \pm 0.030 \unit{R_{\sun}}$       \\
$M_*$                   & $1.353^{+0.080}_{-0.079} \unit{M_\sun}$      \\
$T_{\mathrm{eff},*}$                   & $6460 \pm 140 \unit{K}$        \\
Spectral Type                   & F6 V              \\
\hline
\multicolumn{2}{c}{\hspace{-0.3cm} WASP-121 b}           \\
\hline
$P$                   & $1.27492550^{+2.0 \times 10^{-7}}_{-2.5 \times 10^{-7}} \unit{days}$           \\
$R_P$                   & $1.865 \pm 0.044 \unit{R_J}$           \\
$M_P$                   & $1.183^{+0.064}_{-0.062} \unit{M_J}$          \\
$a$                   & $0.02544^{+0.00047}_{-0.00050} \unit{au}$             \\
$g$                   & $8.43 \unit{m/s^2}$        \\
$T_{\mathrm{eq}}$                   & $2358 \pm 52 \unit{K}$              \\
$T_{\mathrm{irr}}$                   & $3334 \unit{K}$         
\enddata
\tablenotetext{}{All values come from \cite{delrez_wasp-121_2016} except for $g$ and $T_{\mathrm{irr}}$ which we calculated from the others for the GCM models.}
\end{deluxetable}

\begin{deluxetable*}{lccccc}
\tablecaption{Summary of the GCM Suite \label{tab:gcmpres}}
\tabletypesize{\scriptsize}
\tablehead{\colhead{Descriptive Model Name}   &  \colhead{Short Name}   &  \colhead{GCM Code}                                  & \colhead{Applied Drag} & \colhead{Other Physics} & \colhead{Source}}                        
\startdata
Base picket fence & PfBase & RM-GCM & none & none & (1) \\
Cloudy picket fence & PfCld & RM-GCM & none & Active clouds & (1) \\
Magnetic picket fence & PfMag & RM-GCM & Magnetic & none & (1) \\
Cloudy and magnetic picket fence & PfCldMag & RM-GCM & Magnetic & Active clouds & (1) \\
Base correlated-k & CkBase & SPARC/MITgcm & none & none & (2) \\
Correlated-k with H$_2$ dissociation & CkH$_2$ & SPARC/MITgcm& none & H$_2$ dissociation & (3) \\
Correlated-k w/H$_2$ dissoc.\ and weak drag  & CkH$_2$WkDrag & SPARC/MITgcm & Uniform ($\tau_{\rm{drag}} = 10^6$ s) & H$_2$ dissociation & (3) \\
Correlated-k w/H$_2$ dissoc.\ and strong drag  & CkH$_2$StDrag & SPARC/MITgcm & Uniform ($\tau_{\rm{drag}} = 10^4$ s) & H$_2$ dissociation & (3) \\
\added{Correlated-k with reduced opacities} & CkNoFe & SPARC/MITgcm & none & none & (4) \\
\enddata
\tablenotetext{}{(1) \cite{davenport_analysis_2025} (2). This work (3) \cite{wardenier_phase-resolving_2024} (4). \cite{parmentier_thermal_2018}}
\end{deluxetable*}

A commonality between the model sets from each GCM code is that they contain models of the planet with and without an imposed source of drag. The SPARC/MITgcm models include versions with a uniformly applied Rayleigh drag with timescales of $10^6$ s and $10^4$ s, as described in \ref{sec:sparc}. The RM-GCM models use the kinematic MHD prescription described in Section \ref{sec:rmgcm}, with a chosen global magnetic field strength of $B=3$ G, which results in drag timescales near the photosphere of $\sim10^4$ s  on the dayside and $\sim10^8$ s on the nightside.\footnote{For comparison, the rotation period is $\sim10^5$ s and characteristic dynamical and radiative timescales near the photosphere are $\sim3\times10^4$ s and $\sim10^3$ s, respectively \citep{parmentier_cloudy_2021}.} Both of these drag prescriptions reduce the day-night heat recirculation efficiency, resulting in hotter daysides, cooler nightsides, and a reduction in the offset of the hottest point of the planet from the sub-stellar point \citep{wardenier_phase-resolving_2024,davenport_analysis_2025}. 

The SPARC/MITgcm models include versions with and without H$_2$ dissociation. When present, the effect of H$_2$ dissociation and recombination is to increase the heat transport efficiency of the planet, reducing day-night temperature differences and causing a larger offset of the hottest gas on the dayside from the substellar point \citep{tan_atmospheric_2019}. Among the models with H$_2$ dissociation, there are versions with no, weak, and strong drag. In the dragged models, the drag works to counteract the impact of H$_2$ dissociation, suppressing the day-night circulation and so increasing the day-night temperature contrast and reducing any phase offset \citep{tan_atmospheric_2019}.

The RM-GCM models include versions where clouds are allowed to form, when conditions permit. Due to the hot dayside temperatures on this planet, clouds are mostly confined to the planet's nightside where all species considered except KCl form clouds. While feedback between the clouds and atmospheric thermal structure can result in variability, it is at a negligible level for this planet \citep{kennedy_radiatively_2025}. The presence of clouds on the dayside, near the terminator, is shaped by whether drag is applied to the atmosphere or not. In the cloudy model without drag, the super-rotating equatorial jet shapes the temperature structure such that some clouds can exist in the western region of the dayside, while the model with magnetic drag imposes a more symmetric temperature structure across the dayside, resulting in a more symmetric ring of clouds on the edges of the dayside, as described in more detail by \cite{kennedy_radiatively_2025}. The impact of clouds also presents a similar observational effect to that of drag as they both reduce emission from the nightside, 
\added{but we should expect the spectral signature of clouds to be different from changes to the nightside thermal profile caused by drag; a cloudy nightside should not be degenerate with a cooler nightside if sufficient spectral information is available.}

\subsection{Simulated Spectroscopic Phase Curves \label{sec:postprocess}}

The GCMs predict the three-dimensional atmospheric structure for WASP-121~b, but the radiative transfer within is run at a lower spectral resolution than that of the observations, so from the final atmospheric states predicted by the GCMs, we must calculate higher resolution emission spectra. We use separate methods to do this for the RM-GCM and SPARC/MITgcm models, \added{as the post-processing codes used have been designed for compatibility with their respective GCMs, including consistency in their opacities. While it would be valuable to quantitatively assess the impact of the post-processing code used, a general-purpose 3-D code does not exist and this must therefore wait for future work.}

For the RM-GCM models, we follow the post-processing set-up described most recently in \cite{malsky_modeling_2021} with additional update details provided in Section \ref{gasopac} in the Appendix. We re-grid the GCM output from 50 levels of constant pressure to levels of constant altitude and interpolate to produce 250 vertical layers. A line-of-sight, ray-striking radiative transfer scheme \citep{zhang_constraining_2017} then calculates the emitted- and reflected-light intensity toward the observer, for every location on the planet (in latitude and longitude), at a spectral resolution of $R=10,000$. These calculations are performed at appropriate viewing geometries for 24 orbital phases (evenly sampled in increments of 15$\degree$). These are then combined to generate the spectroscopic phase curves of the planet. The radiative transfer includes opacity from 
the same species that were used to calculate the picket-fence coefficients used in the radiative transfer within the RM-GCM (see Section \ref{sec:rmgcm}). Their mixing ratios are calculated using the \texttt{FastChem} equilibrium chemistry code \citep{stock_fastchem_2018, stock_fastchem_2022}. See Section \ref{gasopac} in the Appendix for further details on the opacities within the post-processing. When the GCM determines they are present, the radiative influence of clouds is included, both in absorption and scattering, as described by \cite{malsky_modeling_2021}. The reflected starlight component of the spectrum assumes that stellar flux is a blackbody calculated with the stellar parameters for WASP-121 (Table \ref{tab:stellarparam}). The same cloud species are used in both the post-processing and the GCM runs, and the condensation curves, vertical mixing extents, and optical properties of the species are also consistent between the post-processing and the GCM. While the reflected light component of the generated planetary spectrum self-consistently matches the distribution of clouds in the model, it neglects the small Bond albedo ($A_B=$0.05) applied to the top of the model atmosphere, which is meant to capture the reflection from Rayleigh scattering \citep{malsky_direct_2024}. \added{The assumption of a blackbody over a stellar spectrum for reflection from clouds has minor impact on the results found in this work as clouds are very limited on the dayside, the reflection is only appreciable at the shortest wavelengths, and the primary difference between the blackbody and stellar spectrum are the presence of stellar lines (which are clearly different from the planetary lines).}

For the SPARC/MITgcm models, spectra are calculated \added{with the \texttt{PICASO} open-source radiative transfer code \citep{batalha_exoplanet_2019, mukherjee_picaso_2023}, which was developed from the radiative transfer code in \cite{marley_thermal_1999} and so shares the same heritage as the SPARC/MITgcm.} 
\added{The sources of opacity within the GCM for the CkNoFe model from \cite{parmentier_thermal_2018} originate from \cite{freedman_gaseous_2014}, also matching those used in the RM-GCM. The four other SPARC/MIT models, and the post-processing of all five, used an updated opacity database from \cite{marley_sonora_2021}. The models share the same opacity sources as the CkNoFe model as well as additional sources, the most notable of these is Fe, which causes a slight decrease in the temperature of the dayside photosphere. For a full list of the specific opacity sources used in these models see \cite{tan_modelling_2024}.}
All SPARC/MITgcm models have emission spectra calculated at 36 orbital phases (evenly sampled in increments of 10$\degree$) with a resolution of 622 bins from 0.26 to 267$\unit{\mu m}$ and do not include reflected light in their spectra.

\subsection{Generating Phase Curves and Offsets \label{sec:metoff}}

The JWST phase curve data are observed as the planet-to-star flux ratio ($F_p/F_*$), as a function of wavelength and changing with time. To calculate $F_p/F_*$ for the RM-GCM models we divide the planet fluxes predicted from each model by a PHOENIX spectrum (T$_{\rm{eff}}$ = 6500 K, log(g(cm$^2$/s)) = 4.5, [Fe/H] = 0.0) that is the best match for the WASP-121 star \citep{husser_new_2013} and then convolve the spectrum down to R$=250$ for comparison with the observations. In Figure \ref{fig:allrm} we show the $F_p/F_*$ spectra from each model, at eight evenly-sampled phases throughout the planet's orbit. \added{It is important to note that while the planetary and stellar radii used as inputs in the GCMs are from \cite{delrez_wasp-121_2016} ($R_P/R_*$ = 0.131), the $F_p/F_*$ spectra were calculated using the planet-to-star radius ratio found with NIRISS Order 1 in \cite{Splinter2025} ($R_P/R_*$ = 0.122). If we were to use the radius ratio from \cite{delrez_wasp-121_2016}, the $F_p/F_*$ values would increase by 17\%; see Section \ref{sec:starlight} for further discussion on this and the choice of radius.}

For all models the sets of $F_p/F_*$ spectra at different planetary phases are used to generate white-light phase curves and calculate offsets of the brightest longitude from the sub-stellar point. For the white-light phase curve, we integrated across the whole NIRISS wavelength range of 0.6 - 2.8$\unit{\mu m}$. We also binned the phase curves more narrowly in wavelength, over bin widths of 0.023$\unit{\mu m}$ for the RM-GCM models and 0.03$\unit{\mu m}$ for the SPARC/MITgcm models. This discrepancy in binning between models does not impact our comparison with the data, as in all cases this sufficiently resolves the signatures of any spectral features present. To generate the wavelength-binned phase curves we integrated each simulated spectrum across the chosen wavelength range and then, using the CubicSpline function from \texttt{SciPy} \citep{virtanen_scipy_2020}, interpolated these values across the orbital phases in order to provide higher precision when determining the offset of the peak emission from the time of secondary eclipse (at an orbital phase of 0.5, but the eclipse itself was not modeled). \added{While the SPARC/MIT and RM-GCM outputs have different number of phases, we tested our phase curve analysis at an artificially lowered phase resolution and found only a minor impact on the calculated wavelength-dependent offsets, for instance at a significantly lower level than the differences between models with different choices in drag strength.} 

\section{Results} \label{sec:results}

The NIRISS spectroscopic phase curve of WASP-121~b is an information-rich dataset and this lets us compare our 3-D models to the data in different ways. First, in Section \ref{sec:broadband}, we compare the models to the white-light (wavelength-integrated) phase curve to evaluate: the overall energy budget of the planet, the day-night flux differences, and whether the circulation pattern significantly shifts the temperature structure away from the irradiation pattern or not (i.e., whether there is an offset in the peak of the phase curve from secondary eclipse). After this evaluation of the global energetics, we then investigate what we can learn from the emission spectra of the planet, observed at various phases, about the temperature structure and sources of opacity (including clouds) in those regions (Section \ref{sec:daynight}). Finally, we analyze the wavelength-dependent phase curve offsets to probe how the circulation pattern influences different layers of the atmosphere (Section \ref{sec:resoffsets}).

\subsection{White-light Phase Curves \label{sec:broadband}}

\begin{figure*}
    \centering
    \includegraphics[width=0.8\linewidth]{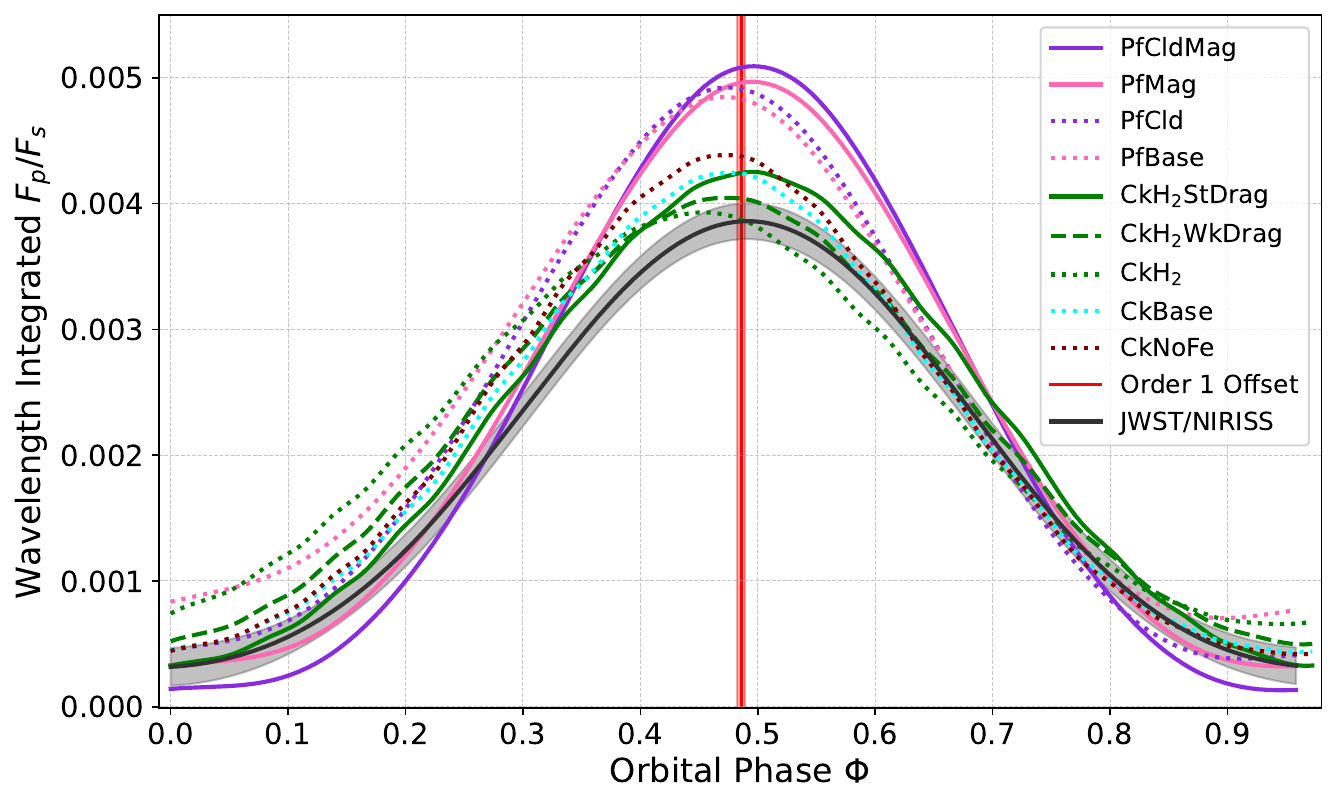}
    \caption{The best fit to the observed white-light phase curve of WASP-121~b from \cite{Splinter2025}, integrated over both the Order 1 and Order 2 NIRISS wavelengths, 0.6-2.85$\unit{\mu m}$ (black line, with gray uncertainty), compared to the equivalent phase curves predicted by our set of 3-D models for this planet (see Table \ref{tab:gcmpres} for more detail). The secondary eclipse occurs at an orbital phase of 0.5, when the dayside of the planet would otherwise be fully in view. The models that most closely match the total emission from the planet use correlated-k radiative transfer. 
    The models that most closely match the observed alignment between the brightest region of the planet and the substellar point \citep[Order 1: 5.1$\pm$1.2$\degree$, Order 2: 10.4$\pm$9.6$\degree$;][]{Splinter2025} are those models with a form of strong drag (solid lines).}
    \label{fig:phasecurve}
\end{figure*}

Figure \ref{fig:phasecurve} shows the white-light (wavelength-integrated over both Order 1 and Order 2) NIRISS/SOSS phase curve of WASP-121~b, compared with the white-light phase curves predicted by the GCMs. \added{The main difference between the data and the models is that most models over-predict the emission from the planet: the models with picket fence radiative transfer are the most discrepant with globally integrated emission 15\% higher than the data, on average, and a significant over-estimation of dayside flux. The models that use a correlated-k scheme are 12\% higher on average and have less egregious overestimation of the dayside flux.} 

We use the NIRISS white-light phase curves to constrain the global energy balance of the planet \added{through comparison with the 3-D models. This is complementary to the Energy Balance Model analysis \cite{Splinter2025} applied to these data and the analysis of the dayside spectrum by \cite{Pelletier2025}, who tested fits that included a reflected light component. Both of these analyses found evidence for significant reflection of starlight from WASP-121~b, characterized by a Bond albedo of $A_B=0.277\pm0.016$ \citep{Splinter2025} and a geometric albedo of $A_g=0.16\pm0.02$ \citep{Pelletier2025}, which is also similar to albedos measured from TESS observations of this planet \citep{wong_systematic_2020,daylan_tess_2021}. Our comparison to the 3-D models finds that their relatively negligible Bond albedos (<0.05) over-estimate the flux observed in the emission-dominated NIRISS range by $\sim$10-15\%, indirectly implying a higher albedo for the planet.}\footnote{We remind the reader that the NIRISS phase curves shown in Figure \ref{fig:phasecurve} cannot be integrated by-eye to estimate the full global emission from the planet. In particular, the models with the coolest night-sides (e.g., PfCldMag and CkH$_2$StDrag) will be emitting more at longer wavelengths than the NIRISS bandpass \citep[see][]{Splinter2025}} 
\added{We discuss the complexity of correctly modeling the planet's absorption of starlight and global thermal emission in Section \ref{sec:starlight}, including why the models with correlated-k radiative transfer more reliably capture this planet's thermal state than the picket fence models. The impact of differing assumptions for atmospheric opacity sources can be seen in the comparison between CkBase and CkNoFe models; the CkBase model, which includes additional opacity sources such as Fe, has global thermal emission 1.0016 times higher than CkNoFe. But, in the NIRISS wavelength range, CkBase has a lower dayside flux than CkNoFe due to the absorption of stellar flux being at lower pressures in the atmosphere, thus reducing the temperature in the deeper photosphere (as can be seen in Figure \ref{fig:day-nightPTs}).}

CkH$_2$ best matches the emission level of the observed dayside because the dampening effect of H$_2$ dissociation sufficiently lowers the dayside emission from the model's overestimation of the global emission. However, this model does poorly in matching other properties of the observed phase curve: it, as is expected, predicts a brighter nightside than observed and shows an eastward offset in the brightest region of the planet away from the substellar point, whereas in the observations the brightest region is well aligned with the substellar point. This observed alignment is best matched by models that include a strong source of drag in the atmosphere (PfMag, PfCldMag, or CkH$_2$StDrag); we further explore how models that incorporate drag best match the NIRISS offsets in Section \ref{sec:resoffsets} and the observed spectra at symmetric phases in Section \ref{sec:sym}. These are also the models whose nightside emission is closer to the observed value (and, as explored further in Section \ref{sec:nightside}, the nightside spectrum as well).

The model with weak drag (CkH$_2$WkDrag) provides a useful reference for comparison between the multifaceted effects that shape white-light phase curves. It has an offset and nightside emission intermediate between the otherwise-identical models with no- or strong-drag (CkH$_2$ and CkH$_2$StDrag) and it has an offset similar to the CkBase model but a brighter nightside. The CkBase and CkH$_2$WkDrag models have two differences in their physical assumptions: the presence of some (weak) drag and the influence of H$_2$ dissociation. Each of those works differently to influence the overall heating of the planet and the day-night heat transport, and present an effective reminder that the day-night temperature difference and phase offset are shaped distinctly by different components of the circulation pattern \citep{roth_hot_2024}.

\added{The lack of a significant offset in the peak of the NIRISS white-light phase curve indicates that strong drag is shaping the circulation of WASP-121~b; this is consistent with previous results for the planet \citep{bourrier_optical_2020,daylan_tess_2021,mikal-evans_diurnal_2022,mikal-evans_jwst_2023,morello_spitzer_2023,davenport_analysis_2025} and results for other UHJs: WASP-76~b \citep{beltz_exploring_2022}, WASP-103~b \citep{Kreidberg2018}, WASP-18~b \citep{coulombe_broadband_2023}. While the strong drag } limits heat transport to the nightside and so decreases emission at those phases, we can see that the presence or absence of clouds also impacts the phase curve. When nightside conditions are cool enough for clouds to form ($\lesssim1500$ K, and the cloudy models predict that this should be the case), then the clouds can blanket the thermal emission from the nightside, making it appear even dimmer. These effects can combine, as seen in the model with both clouds and magnetic drag, PfCldMag, which has the dimmest nightside of all the models. Here the model with clouds but no drag (PfCld) is an interesting comparison case, as it predicts a dim nightside in rough agreement with the data, but has a peak offset inconsistent with the observations (see also Section \ref{sec:resoffsets} and Figure \ref{fig:alloffset}), as it is lacking a strong source of drag. Again, we see that multiple physical effects can shape the atmospheres of UHJs, but by considering all of the evidence contained within a phase curve, we can more confidently interpret the causes of the observed features. In Section \ref{sec:daynight} we more closely examine the emission from the nightside, including the additional insight we gain from the spectral information.

\subsection{Spectra at Different Phases}  \label{sec:daynight}

Thanks to the spectroscopic nature of the NIRISS/SOSS data, we can also extract the planet's emission spectrum at different phases, recovering the hemispherically integrated flux from the dayside, nightside, and mixtures of the two, providing further insight into how the atmospheric circulation on this planet shapes its 3-D temperature and chemical structure.

\subsubsection{Dayside Emission Spectrum} \label{sec:dayside}

\begin{figure*}[t!]
\begin{center}
\includegraphics[width=1\textwidth]{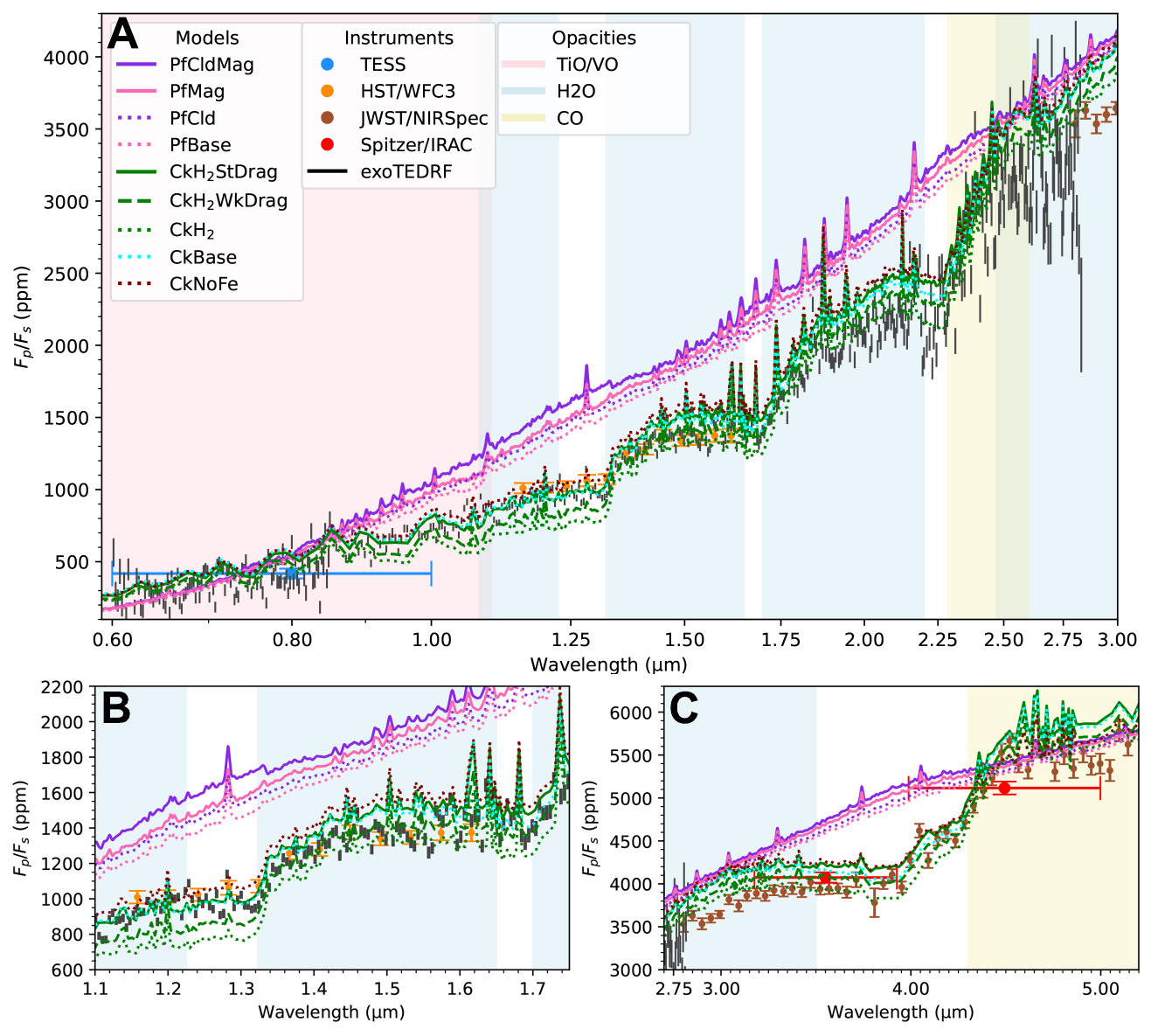}
\end{center}
\vspace{-0.8cm}
\caption{The dayside spectrum of WASP-121 b as observed by JWST/NIRISS (black lines with errors) compared to photometric and spectral measurements from other instruments (referenced in the main text) and simulated dayside spectra from our set of 3-D models (see Table \ref{tab:gcmpres} for more detail). Spectral features from some expected sources of opacity occur in the shaded wavelength regions. \added{Panel A shows the observed and simulated spectra for the range observed by NIRISS/SOSS (0.6-2.8 $\unit{\mu m}$). Panel B focuses on the area observed both by NIRISS and HST (1.12 - 1.65 $\unit{\mu m}$) and highlights a prominent water feature. Panel C extends past the NIRISS range and shows the models compared to Spitzer and NIRSpec (2.70-5.5 $\unit{\mu m}$) observations. The models that use correlated-k radiative transfer most closely match the NIRISS 
and other data. CkH$_2$WkDrag and CkH$_2$StDrag, which incorporate both drag and H$_2$ dissociation, and CkBase, which incorporates neither of these, are the best fits for the data, although they generally overestimate the flux at longer wavelengths.}} 
\label{fig:allday}
\end{figure*}

\added{\cite{Pelletier2025} presented a retrieval analysis of this NIRISS dayside spectrum, finding that it is best fit by a metal-enriched, inverted atmosphere \citep[as previously identified by][]{evans_ultrahot_2017} with the presence of H$_2$O, VO, CO, and H$^-$, and either TiO or a reflected light component. Here we compare NIRISS dayside data against our 3-D models.} In Figure \ref{fig:allday} we
include both the NIRISS spectrum, as well as previous observations of WASP-121~b's dayside from TESS \citep{daylan_tess_2021}, HST/WFC3 \citep{mikal-evans_diurnal_2022}, Spitzer \citep{davenport_analysis_2025}, and JWST/NIRSpec \citep{Evans-Soma2025}. 
\added{We see that while the overall emission from the correlated-k 3-D models is slightly too high, as seen with the white-light curve analysis, they do reproduce spectral features in emission.} 
In contrast, the picket fence models have dayside spectra that are too bright and too isothermal \added{(see also the dayside pressure-temperature profiles in Figure \ref{fig:day-nightPTs})}.\footnote{The small, sharp apparent emission features in these $F_p/F_s$ spectra are actually stellar absorption lines that become flipped when the planetary spectra are normalized by $F_s$.} We discuss this limitation of the picket fence approach further in Section \ref{sec:picket-fence}.

\added{The models that best fit the NIRISS dayside data are CkBase, CkH$_2$WkDrag, and CkH$_2$StDrag models. They have reduced $\chi^2$ values\footnote{In this case we are just reducing $\chi^2$ by the number of data points, as there is no clearcut way to count the number of parameters in the GCMs.} of, respectively, 14.1, 13.2, and 15.5. This highlights the complexity of competing physical effects; the addition of $H_2$ dissociation decreases the dayside emission but an imposed drag can, effectively, counteract this effect. In addition, while these models are the best fit for the NIRISS data, it can be seen in Panel C of Figure \ref{fig:allday} that CkH$_2$ best fits the NIRSpec data\footnote{The star-to-planet radius ratio assumed for the post-processing was chosen due to its alignment with the NIRISS wavelength range, and so can be misrepresenting the emission at these longer wavelengths. See Section \ref{sec:starlight} for further details.} and, as discussed earlier, best matches the total emission of the dayside.}

\subsubsection{Nightside Emission Spectrum} \label{sec:nightside}
\begin{figure*}[t!]
\begin{center}
\includegraphics[width=1\textwidth]{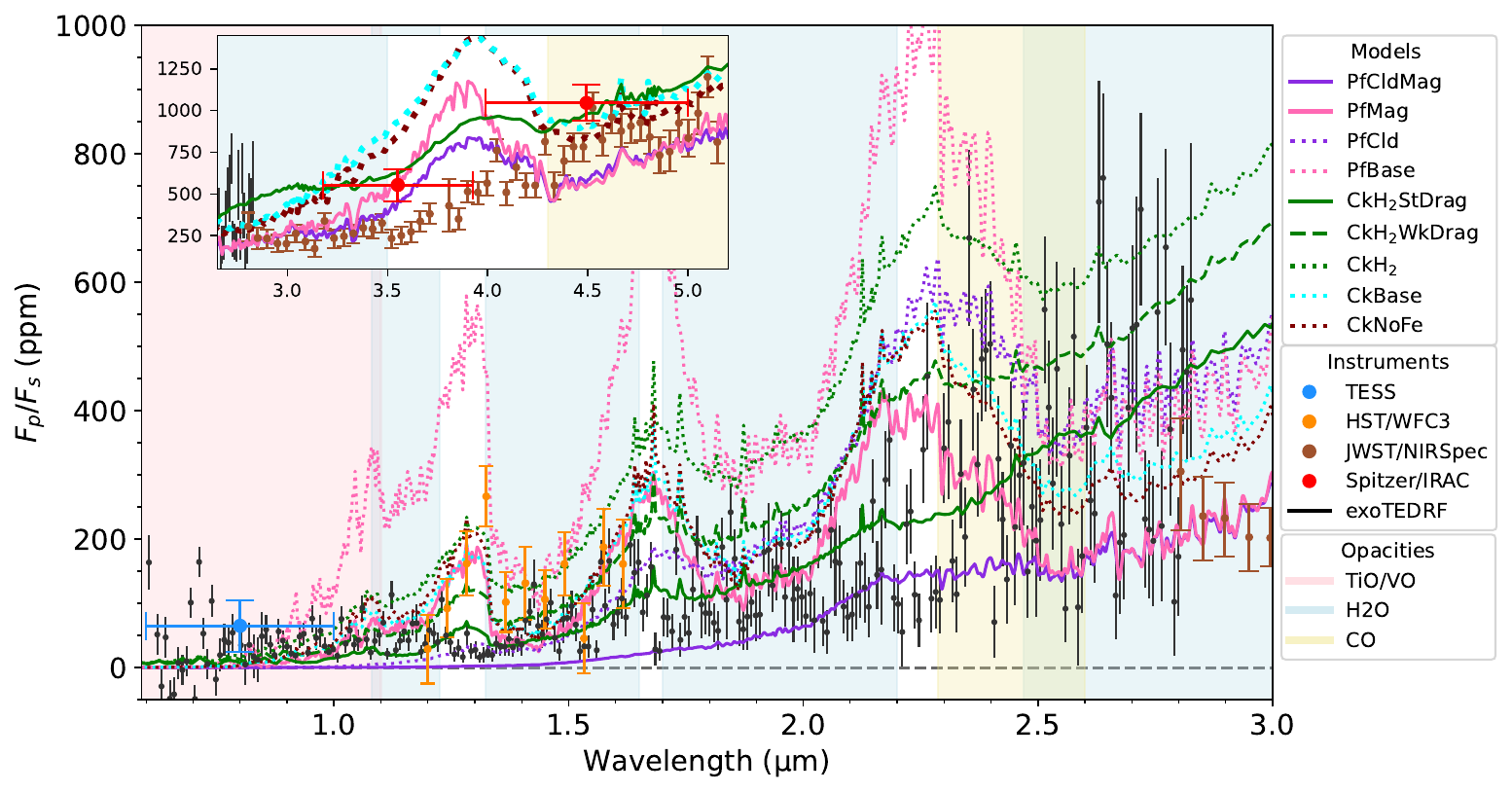}
\end{center}
\vspace{-0.8cm}
\caption{The nightside emission spectrum of WASP-121~b as observed by JWST/NIRISS (black points with errors) compared to photometric and spectral measurements from other instruments (referenced in the main text) and simulated nightside spectra from our set of 3-D models (see Table \ref{tab:gcmpres} for more detail). Spectral features from some expected sources of opacity occur in the shaded wavelength regions. The models that most closely match the NIRISS data (and generally the other data) are those that have a lower level of nightside flux and are relatively featureless due to nightside clouds (PfCldMag, purple line) or the hemispheric averaging of inverted and non-inverted regions (CkH$_2$StDrag, green line).}
\label{fig:allnight}
\end{figure*}

Figure \ref{fig:allnight}, in which we compare the 3-D models with the observed nightside spectrum (including data from other instruments, from the same references given above), shows that while the CkBase and CkH$_2$WkDrag models are good matches to the dayside, they are not in fact a good match to the whole planet. As indicated by the white-light phase curve and confirmed here, the NIRISS data are in best agreement with those models that have the dimmest nightsides. While noisy, the NIRISS spectrum also shows that the nightside emission from this planet is not just dim but also relatively featureless. Note that, while the dayside spectra from NIRISS and HST/WFC3 were in good agreement, for the nightside we see a discrepancy between the two, with the HST data finding a strong water absorption feature \citep{mikal-evans_diurnal_2022} that is not seen in the NIRISS data. While variable emission from the planet could be a potential explanation \citep{changeat_2024}, we do not see evidence of variability from other instruments or at other phases. \added{None of our models predict significant variability, including the cloudy ones \citep{kennedy_radiatively_2025}, although other hot Jupiter GCMs with clouds have sometimes found variable and/or patchy nightside clouds \citep{Lines2018,komacek_patchy_2022}.}

\begin{figure*}
    \begin{center}
        \includegraphics[width=0.9\textwidth]{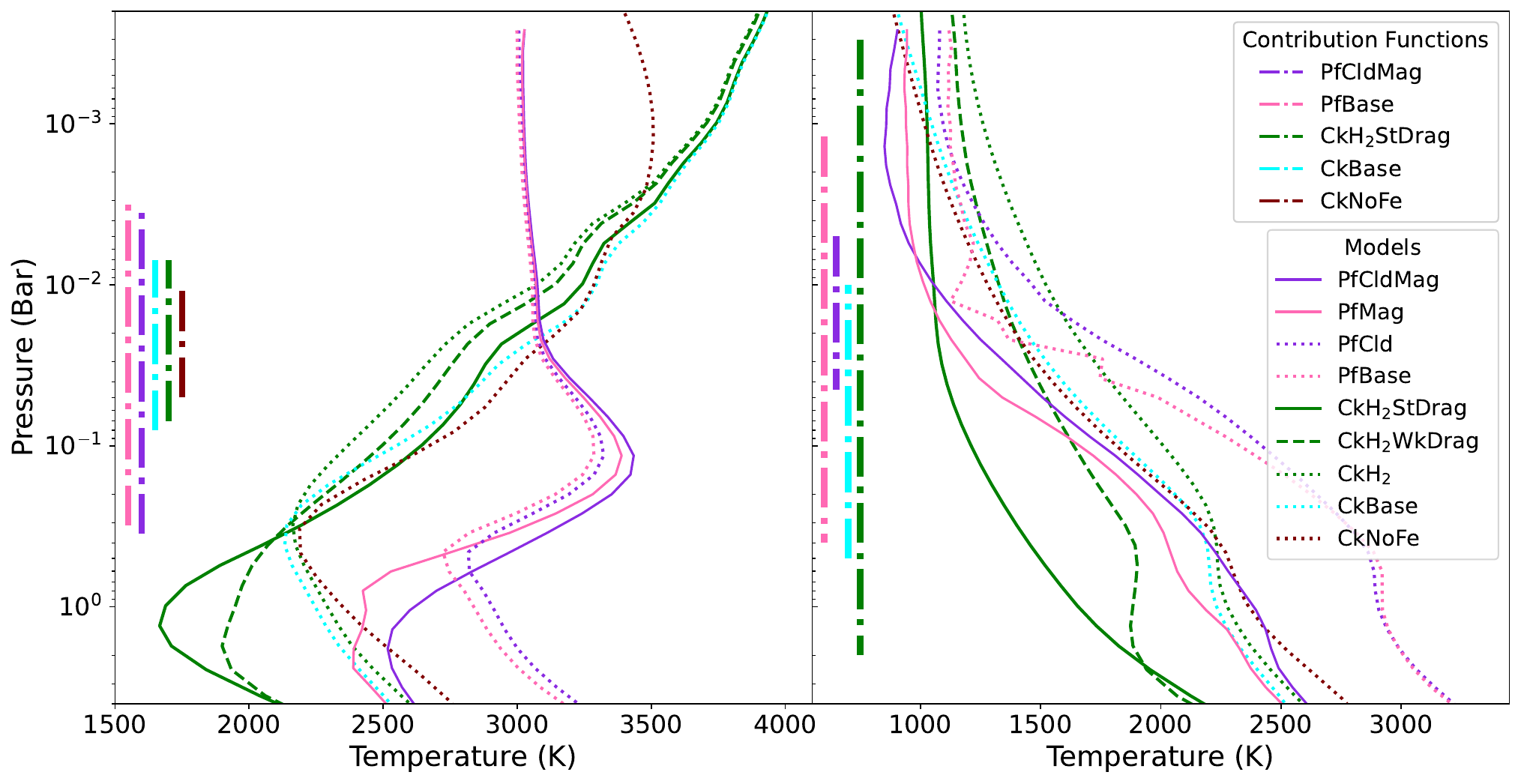}
    \end{center}
    \vspace{-0.7cm}
    \caption{Spatially averaged pressure-temperature profiles of the 3-D models of WASP-121~b; on the left, the dayside, and the right, the nightside. The profiles are averaged over a 30$\degree$ region around the substellar and antistellar points for, respectively, the day and nightside. The vertical lines on the left-hand side indicate pressure ranges where observed emission originates from. On the dayside, the correlated-k models produce similar atmospheric structures, all with strong temperature inversions. On the nightside, the emission from the planet is strong influenced by the presence or absence of clouds (shifting the contribution function). The recombination of H$_2$ heats the nightside, strong drag keeps the nightside cooler, and when both effects are included the nightside profiles have smaller temperature gradients (i.e., are more isothermal).}
    \label{fig:day-nightPTs}
\end{figure*}

The NIRISS data do not show evidence of the strong absorption features predicted by many of the 3-D models and a simple $\chi^2$ comparison confirms that the models with muted nightside spectra fit the data better than models with nightside spectra that are dim but include spectral features. The PfCldMag and CkH$_2$StDrag models have reduced $\chi^2$ values of 3.6 and 3.8, respectively, while the PfMag and CkBase models have respective values of 9.7 and 13.0. \added{We can diagnose the causes of the muted nightside spectra predicted by some models in Figure \ref{fig:day-nightPTs}, where we show pressure-temperature profiles from the 3-D models, averaged around the anti-stellar point, and also the pressure ranges over which we receive emission from each model. For the PfCldMag model, the clouds raise the photosphere to be higher in the atmosphere and the emitted photons come from a narrower range of pressure, minimizing the temperature gradients probed and muting the spectral features, especially at wavelengths shorter than 2$\unit{\mu m}$. For the CkH$_2$StDrag model, the reason for the lack of spectral features is a little more complex. \cite{tan_modelling_2024} discusses how the nightside emission from the CkH$_2$StDrag model includes contribution from both the broad cool antistellar photosphere at $\sim$1 bar (seen in Figure \ref{fig:day-nightPTs})) and the very hot and thermally inverted near-terminator regions (shown in Figure \ref{fig:symmetry}). 
We can also see in Figure \ref{fig:day-nightPTs} that the presence of H$_2$ recombination (and associated heating) together with strong drag works to create more isothermal conditions near the anti-stellar point, compared to less dragged models and ones without H$_2$ recombination}.

It may be possible to differentiate between a cloudy nightside and one shaped by a combination of H$_2$ recombination and strong drag by obtaining more observations at shorter or longer wavelengths. With higher signal-to-noise observations in the optical, we would expect to see contributions from just the terminator regions in the CkH$_2$StDrag model, resulting in strong spectral features in emission \citep{tan_modelling_2024}, while for the PfCldMag model we would still expect to see strongly muted spectral features. At longer wavelengths, the Spitzer and JWST/NIRSpec data seem to prefer the cloudy scenario and with even more extended wavelength coverage from JWST/MIRI we would, based on the nightside profile of PfCldMag, expect to observe a silicate cloud resonance feature around 10$\unit{\mu m}$ \citep{inglis_quartz_2024}. However, the MIRI phase curve of another hot Jupiter, WASP-43~b, offers a cautionary tale: while it implies a cloudy nightside, it does not show any spectral features from clouds \citep{bell_nightside_2024}, meaning that this may not always be as obvious of a signature as we might hope.

\subsubsection{Emission Spectra from Symmetric Phases} \label{sec:sym}

\begin{figure*}[t!]
    \centering
    \includegraphics[width=0.8\linewidth]{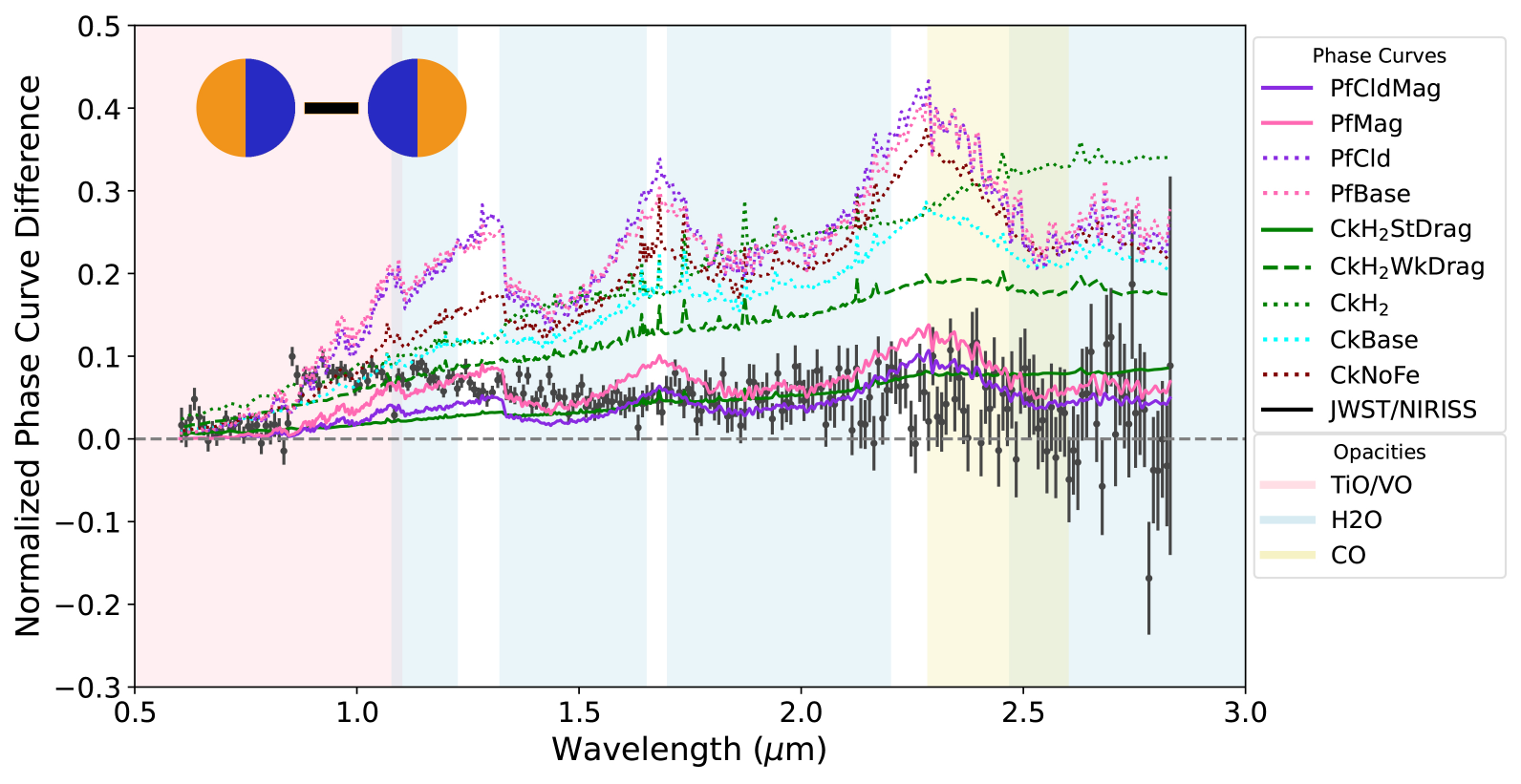}
    \caption{The difference between the $F_p/F_s$ NIRISS spectra (black points) of WASP-121~b at symmetric phases (relative to day/night) of $\pm90\degree$, normalized by the maximum $F_p/F_s$ value in each pair of spectra. These are compared to the simulated spectra from our set of 3-D models (colored lines). Shaded regions indicate important sources of opacity within UHJ atmospheres. At $\pm90\degree$ the day and nightside of the planet are equally in view, but opposite sides of the planet are seen. The observed symmetry between these phases are best matched by models with a strong source of drag applied (solid lines). The observed and modeled spectra at each phase, before differencing and normalization, are included in Section \ref{sec:appendplot} in the Appendix.
    \label{fig:symdiff}}
\end{figure*}

In the Appendix (Section \ref{sec:appendplot}) we show the NIRISS spectra observed at eight equally spaced phases throughout WASP-121~b's orbit, as well as the equivalent spectra predicted from each 3-D model. Aside from the mismatch between predicted and observed flux levels, already discussed in Section \ref{sec:broadband} above, and the transition from spectral features in emission on the dayside to being muted on the nightside (shown in Figures \ref{fig:allday} and \ref{fig:allnight}), the other interesting result is the fairly close match between spectra from symmetric phases. 
We quantify this in Figure \ref{fig:symdiff}, which shows the difference between spectra from the phases of $\pm90\degree$. These phases are equally spaced before and after eclipse (or transit) and show equal parts of dayside and nightside of the planet.
If both the dayside and nightside were completely east-west symmetric, then any observed difference would be 0. For a planet with primarily eastward circulation (as is typical for hot Jupiters), the regions to the east of the substellar point are brighter than those to the west and, on the nightside, the regions to the west of the anti-stellar point are brighter than those to the east of it (due to advection of hot gas from the dayside). The differences between phases in Figure \ref{fig:symdiff} have been taken such that positive values correspond to a brightness pattern that matches predominantly eastward advection. 
We see that models with little to no drag have stronger eastward asymmetry in their emission than is observed in the NIRISS data, in agreement with the behavior shown in the white-light phase curve (Figure \ref{fig:phasecurve}).

The models that best match the phase-symmetry in the data are those with a form of strong drag (CkH$_2$StDrag, PfCldMag, and PfMag), providing another piece of evidence to showcase the central role that strong drag has in shaping the atmospheric structure of WASP-121~b. There is, however, a mismatch between the symmetry predicted by strong-drag models and the observed spectroscopic phase curve at wavelengths between $\sim0.8-1.2$ $\mu$m, indicating more eastward advection at the pressures probed by those wavelengths than is predicted by the strong-drag models. 

\added{The spectral difference measurements between symmetric phases become increasingly noisy past $\sim$1.5 $\mu$m, lacking the precision necessary to differentiate between a noticeable distinction between the strongly dragged picket fence and correlated-k models: whether the phase asymmetries are fairly uniform in wavelength or show spectral features. More generally, in Figure \ref{fig:symdiff} we see that it is the models that include H$_2$ dissociation and recombination that exhibit spectrally uniform differences in phase, while all models that lack this physical effect have larger differences at continuum wavelengths than in spectral features. This can be seen best in comparing CkH$_2$ versus CkBase, where their phase curve differences are equal when in continuum wavelengths, but diverge when in spectral features. We show averaged pressure-temperature profiles from each model, at $\pm90$ degrees from the substellar point, in the Appendix (\ref{sec:appendTPs}). A careful evaluation of these thermal profiles and the spectroscopic phase curves points toward there being a more uniform change in temperature, across the pressures probed by the continuum and spectral features, between symmetric regions of the planet in the models with H$_2$ dissociation than those without. However, we do not identify this as a straightforward way to search for the influence of H$_2$ dissociation, as the spectral uniformity of phase curve differences depends on a complicated combination of changes in thermal profiles and contribution functions, within a fully 3-D context.}

Comparisons between models and observations across all phases of the planet's orbit demonstrate how the three-dimensionality of UHJs truly requires global measurements of a planet, if we hope to accurately interpret the data. For example, although the white-light phase curve and symmetric phases indicate the presence of strong drag in this atmosphere, the PfMag model shows that strong drag alone is not sufficient to produce the featureless nightside spectrum observed. The PfMag model also predicts a nightside spectrum that is\added{, until around 2.2 $\unit{\mu m}$,} strangely well aligned with that of the CkBase model, which has a substantially different global energy balance, uses completely different radiative transfer, and has no drag applied. While these models coincidentally arrive at similar nightside temperature continuum levels, they are not equally good representations of the planet. We also see the importance of understanding UHJs as 3-D objects in the seeming dichotomy implied by the day- and night-side spectra: the models that best match the dayside are clear, \added{weak or drag-free, and only one has H$_2$ dissociation,} while the models that best match the nightside and symmetry between phases have strong drag and either clouds or H$_2$ dissociation.

\subsection{Wavelength-dependent Phase Curve Offsets \label{sec:resoffsets}}

From the alignment of the peak in the white-light NIRISS phase curve with secondary eclipse (Figure \ref{fig:phasecurve}), we already know that the brightest region of the atmosphere is near the substellar point, indicating a source of strong drag in the atmosphere. We can also use the spectroscopic nature of the NIRISS data to look for differences in phase curve offset with wavelength, as we will see emission from different depths into the atmosphere when we are in or out of opacity bands and so can obtain more information about the planet's 3-D circulation pattern \citep[e.g.,][]{dobbs-dixon_wavelength_2017,parmentier_cloudy_2021}. 
In particular, this may help us to differentiate which form of drag to expect (uniform or magnetic); the highest levels of thermal ionization occur in the hot, low-density regions of the upper dayside, meaning that the influence of magnetic drag should become weaker in deeper layers of the atmosphere \citep{perna_magnetic_2010,rauscher_three-dimensional_2013,beltz_exploring_2022}.

\begin{figure*}[t!]
\begin{center}
\includegraphics[width=1\textwidth]{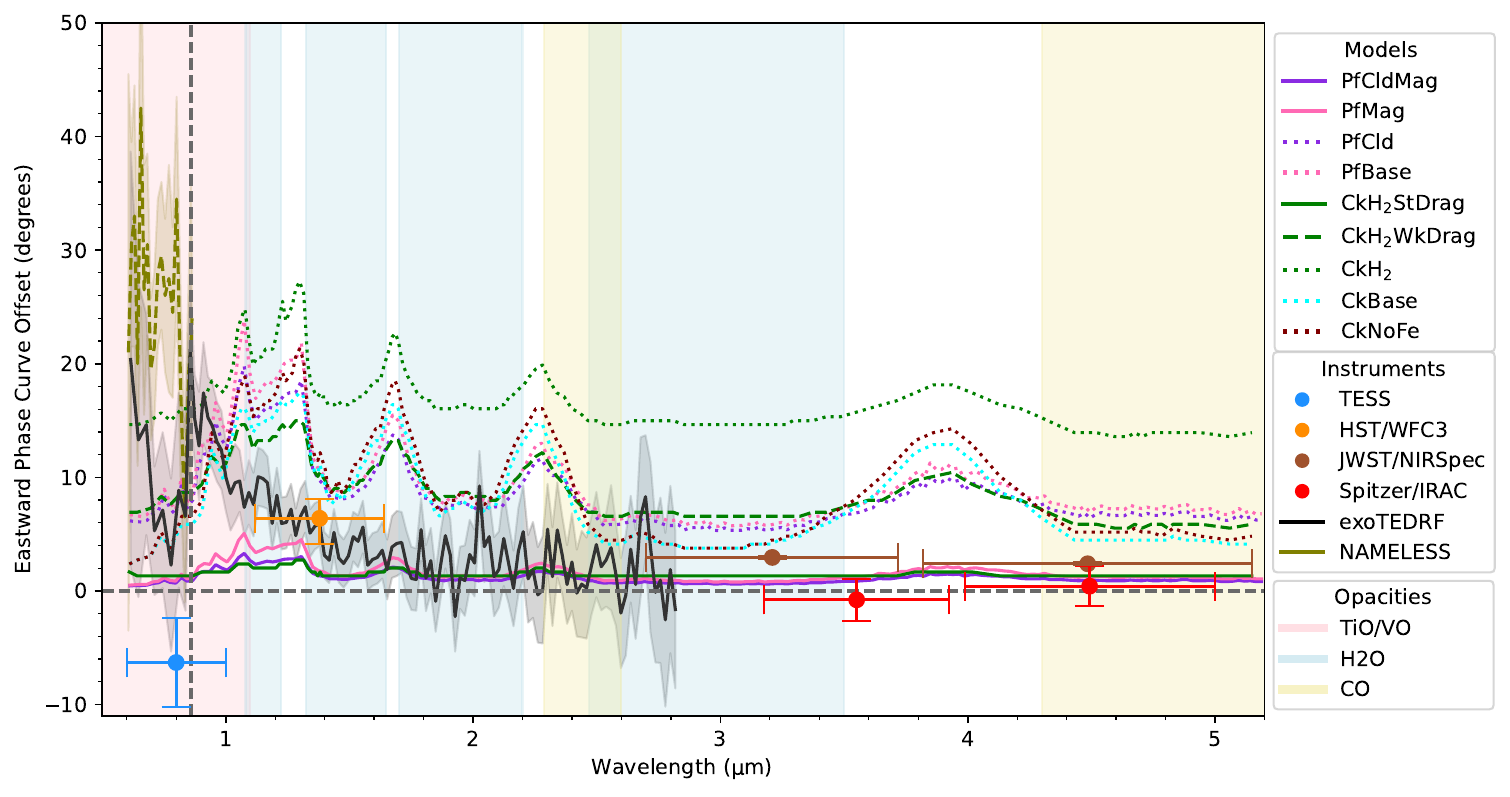}
\end{center}
\vspace{-0.8cm}
\caption{The offset between the brightest region of WASP-121~b and its substellar point, observed as a function of wavelength with JWST/NIRISS, from the exoTEDRF data reduction (black line and gray uncertainties) and, for Order 2 only, the NAMELESS reduction (olive dashed line) as well. The Order 1 NAMELESS reduction is not shown as its offsets are in agreement with exoTEDRF (see Figure C7 in \cite{Splinter2025}). The NIRISS data and other measurements (referenced in the main text) are compared to the wavelength-dependent offsets predicted by our set of 3-D models (see Table \ref{tab:gcmpres} for more detail). The wavelength regions with contributions from common sources of opacity are shaded, as labeled. The break between the Order 1 and Order 2 NIRISS data is marked with a vertical dashed line. The models with strong drag applied (solid lines) best match the data, both in the relatively small offset values and the lack of a strong dependence of offset with being in or out of absorption bands. While data and models are in general agreement long-ward of $\sim$1.4$\unit{\mu m}$, there is significant discrepancy at shorter wavelengths between the NIRISS data, TESS data, and models.}
\label{fig:alloffset}
\end{figure*}

In Figure \ref{fig:alloffset} we show the wavelength-dependence of the alignment between the brightest region of WASP-121~b and its substellar point, as measured by NIRISS and in comparison to previous phase curve observations \citep[from TESS, HST/WFC3, JWST/NIRSpec, and Spitzer/IRAC:][respectively]{daylan_tess_2021,mikal-evans_diurnal_2022,Evans-Soma2025,davenport_analysis_2025} as well as the predictions from our set of 3-D models. Overall there is good qualitative agreement between the phase curve offsets measured by each instrument, although formally the quoted uncertainties do not always agree within 1-$\sigma$. The NIRISS data short-ward of $\sim$1.4 $\mu$m shows a trend of increasing offset with decreasing wavelength that is not well matched by any model but is in the two independent reductions done on the NIRISS data (NAMELESS and exoTEDRF). In addition, there is a change in behavior at the transition between data from Order 1 and 2 (at 0.85$\unit{\mu m}$), although the two reductions differ in this region, and tension with the TESS phase curve offset. These mismatches between expectations and observations are expanded on in Section \ref{sec:waveoffsets}.

Overall, the data continue to favor those models with little-to-no offset. The models with larger offsets are not only inconsistent with the data, but also predict that the offsets should have strong spectral variations. All of the models without strong drag show larger offsets at continuum wavelengths, where we are seeing deeper into the atmosphere (where radiative timescales are longer) and so regions where the winds more efficiently advect hot gas away from the substellar point before it has time to cool. Notably, this even includes the non-dragged picket fence models (PfBase and PfCld); even though those dayside atmospheres are more isothermal than the data indicate, we are still seeing to different enough depths into the atmosphere that the advected temperature structure changes. The strongly dragged models (PfCldMag, PfMag, and CkH$_2$StDrag) also show some dependence of the offset on spectral bands, but at a much weaker level and within the uncertainties of the measurements longward of 1.2 \unit{\mu m}. We do see that the strongly dragged model that uses a uniform drag timescale (CkH$_2$StDrag) shows even less variation in offset with wavelength than the models with magnetic drag (PfCldMag, PfMag), which may be due to the reasons discussed above, but since these models also have different thermal profiles and the magnetic ones lack the influence of H$_2$ dissociation, these effects are difficult to disentangle.

The models with clouds show very little difference in offset from the equivalent clear models (compare PfCldMag to PfMag and PfCld to PfBase), which is unsurprising since the daysides of all these models are predominantly clear and the peak offset is most sensitive to the dayside thermal structure. The cloudy models do show slightly smaller eastward offsets than the clear models and this effect is a bit more pronounced at shorter wavelengths, where there may be some contribution from clouds along the western terminator when drag is not present (Figure \ref{fig:reflectedlight} and discussed more below). The idea here is that the thermal emission may show a slightly eastward shift, from the standard eastward equatorial jet typical of hot Jupiter atmospheric circulation, but this may be slightly counteracted by reflection from the western dayside, resulting in a smaller net eastward shift \citep[e.g.,][]{parmentier_cloudy_2021}. There is the additional effect, applicable to models with and without drag, that the presence of nightside clouds suppresses the flux at every phase besides 0.5 and so will push the offset closer to zero; however,
for the mostly clear daysides here, these effects are smaller than the measurement uncertainties.

\section{Discussion} \label{sec:disc}

In comparing 3-D models to the NIRISS phase curve of WASP-121~b we identified two features in the data that cannot be explained well: 1) the trend in phase offset with wavelength at the shortest wavelengths and 2) the overall level of emission from the planet. Here we discuss each of this issues in more detail, eliminating possible explanations and highlighting remaining uncertainties.

\subsection{Phase Curve Offsets at the Shortest Wavelengths}\label{sec:waveoffsets}

In an attempt to explain the discrepancy between the NIRISS offsets at the shortest wavelengths and predictions from our 3-D models (Figure \ref{fig:alloffset}), we consider two potential mechanisms that could preferentially affect the planetary spectrum at shorter wavelengths: 1) reflected light, especially any contribution from clouds, and 2) the depletion of optical absorbers in the atmosphere. These mechanisms operate at shorter wavelengths and could produce a spatial asymmetry in the light from the planet, as is required to produce an offset.

\begin{figure}[t!]
\begin{center}
\includegraphics[width=1\linewidth]{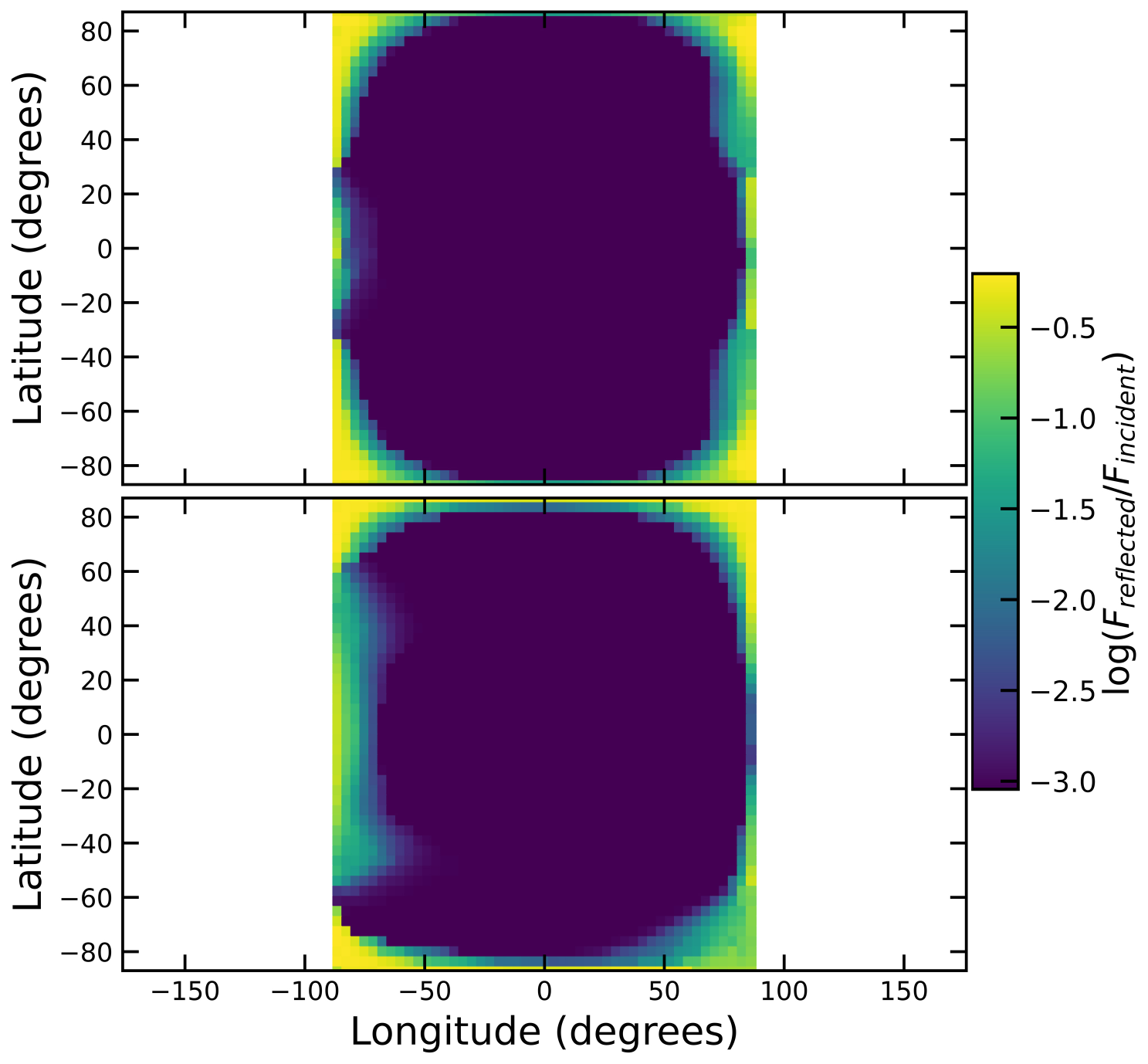}
\end{center}
\vspace{-0.7cm}
\caption{Maps of the ratio between reflected and incident starlight (centered on the substellar point) from the 3-D models that include clouds; top: the PfCldMag model, which includes a prescription for magnetic drag, and bottom: the PfCld model, which does not. 
The centrally located region that dominates the center of the maps are due to gas scattering in the planet's atmosphere. The reflected light pattern at the edges of the dayside is due to clouds. It presents as far more symmetric in the PfCldMag model due to the drag causing symmetry in the dayside temperature structure, while in PfCld the clouds are preferentially condensed on the western part of the dayside due to the eastward advection of cooler air from the nightside.}
\label{fig:reflectedlight}
\end{figure}

We may expect an eastward shift in reflected light from the planet if there are preferentially clouds to the east of the substellar point, leading to enhanced reflection from that region. Two of the models in our suite include a prescription for clouds, but neither of them predicts predominantly eastern clouds, as shown in Figure \ref{fig:reflectedlight} and already indicated by their very small predicted offsets at short (and all) wavelengths (Figure \ref{fig:alloffset}). While these picket fence models have hotter daysides than indicated by the data, if the daysides were cooler we still would not expect clouds to preferentially form to the east of the substellar point, since the standard eastward circulation pattern results in hotter gas to the east of the substellar point (and so fewer clouds). In the presence of strong drag---as indicated by the small phase curve offsets at longer wavelengths---the dayside temperature structure may be more symmetric around the substellar point, but this also does not result in clouds on the eastern dayside. In addition to the specific models for WASP-121 b presented here, more generally we do not expect clouds on the eastern dayside of hot and ultra-hot Jupiters, regardless if drag is present \citep{parmentier_cloudy_2021,kennedy_radiatively_2025}.

The other potential mechanism that could produce an eastward shift at short wavelengths would be if there is a lower atmospheric opacity than expected at those wavelengths, resulting in emission originating from deeper in the atmosphere where the hotter regions are generally more advected to the east. A valuable comparison here is between the picket fence models with the most disparate offsets: PfBase (no clouds or magnetic drag) and PfCldMag (with both effects). The emission from these models is emerging from almost identical pressure ranges within each atmosphere ($\sim$3 to 300 mbar), so the difference in offset between the two is due to their different temperature structures. When we calculate spectra from each model, we can also identify that offsets at each wavelength link very directly with the pressure level probed, so wavelengths with similar offsets have emission from similar pressures. Thus, in agreement with all of the model predictions (both picket fence and correlated-k ones), the way to get increasingly eastward offsets at shorter wavelengths---while maintaining a physically consistent circulation pattern---would be to remove a source of opacity whose absence would then allow for emission from increasingly deep layers.

The gaseous molecules TiO and VO are known to be strong absorbers in the optical, but also could potentially be removed from the atmosphere due to condensation on the nightside \citep{spiegel_can_2009, parmentier_3d_2013}. 

\begin{figure}[t!]
\begin{center}
\includegraphics[width=1\linewidth]{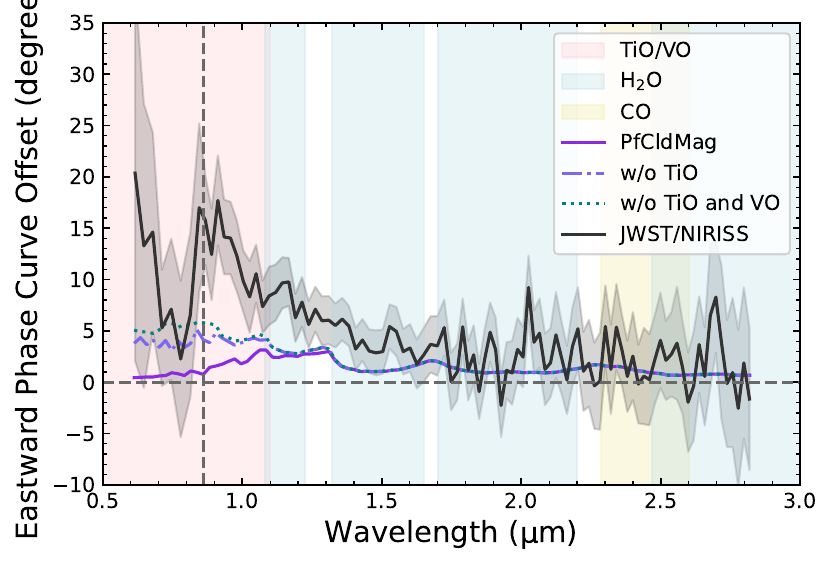}
\end{center}
\vspace{-0.7cm}
\caption{A comparison of the phase curve offsets measured by NIRISS (in black with gray uncertainties) to the values predicted by one of the strongly dragged 3-D models that matches the longer wavelength data well (PfCldMag) and versions of that model where TiO or TiO and VO have been artificially removed when calculating the simulated spectra. The vertical dashed line indicates the transition between Order 1 and Order 2 NIRISS data. While a lack of TiO in the atmosphere (and secondarily VO) would result in emission from deeper regions of the atmosphere with larger eastward offsets, this still does not fully explain the trend seen in the data.}
\label{fig:notio}
\end{figure}

TiO and VO have proved elusive in high-resolution observations of WASP-121 b, but complementary analysis of the NIRISS phase curve dataset have robustly detected VO on the planet's dayside \citep{Pelletier2025} and both TiO, albeit depleted, and VO in the planet's transmission spectrum, probing the terminator region (MacDonald et al. in prep.). In conjunction with these results, we investigate this mechanism from a theoretical standpoint. We recalculated the predicted spectral phase curve from one of the strongly dragged models that otherwise matches the offset data well (PfCldMag) with those species excluded from the post-processing opacities (introducing inconsistency with radiative transfer in the GCM). Figure \ref{fig:notio} shows a comparison between the observations and the offsets predicted by this model with and without TiO and VO. While the removal of these opacity sources does result in emission from deeper in the atmosphere, and so a slightly more eastward offset, it does not reach the magnitude of offset seen in the data nor does it match the slope seen in the NIRISS offsets.

In addition to physical explanations for the trend seen in the NIRISS data, it may also be possible to attribute this signature to the challenge of analyzing phase curve data at the shortest wavelengths. Since the planet's emission drops off at shorter wavelengths, the measurement becomes intrinsically more noisy and sensitive to how any systematics are corrected (as evidenced by the wavelength-binned NIRISS phase curves shown in Figure C4 from \cite{Splinter2025}). In addition, at shorter wavelengths stellar variability can become more of a nuisance, requiring careful correction. \cite{Splinter2025} discuss how correcting for the stellar variations influences the phase curve offsets and find that while the trend of increasing offset with decreasing wavelength seems robust up to the transition from Order 1 to Order 2 (at 0.85 $\mu$m), different choices in the analysis can produce more discrepant results short-ward of that. This can be seen in Figure C7 from \cite{Splinter2025} where the offsets from the independent NAMELESS and exoTEDRF reductions show agreement in Order 1, including the upward trend short-ward of $\sim$1.4 $\mu$m, but diverge in Order 2. The divergence is because only the exoTEDRF reduction accounted for stellar variability in Order 2; this leads to the offsets from the NAMELESS reduction to lack the sharp break between Order 2 and 1 that the exoTEDRF reduction has. Neither reduction aligns with the TESS data, which overlap with this same wavelength region (primarily Order 2 for NIRISS) and also required significant correction for the stellar variability \citep{daylan_tess_2021}.

\subsection{Accurately Modeling Planetary Emission} \label{sec:starlight}

\added{None of the models explored in this work match the global emission levels observed by NIRISS, but this conclusion is complicated by the assumptions in calculating emission spectra, primarily the planet radius used to integrate over the visible hemisphere. There are two important aspects here: 1) the emitting surface of a planet is a function of wavelength and accounting for this in calculating hot Jupiter spectra can result in differences of up to 5-20\% compared to calculations that lack this nuance \citep{Fortney2019}, and 2) it is not obvious how best to ``normalize'' the planet radius, given the first point. 

We addressed the normalization of the planet's radius by, as stated in Section \ref{sec:metoff}, using the planet-to-radius ratio from \cite{Splinter2025} ($R_P/R_*$ = 0.122) which differs from the stellar and planetary parameters used in generating the models \citep[$R_P/R_*$ = 0.131;][also see Table \ref{tab:stellarparam}]{delrez_wasp-121_2016}. We chose to use the values from \cite{Splinter2025} for the $F_p/F_*$ calculations because we are comparing the simulated spectra against the observations from that work and by using values from the same wavelength range we can more accurately capture the appropriate emitting surface of the planet. This is in comparison to the values from \cite{delrez_wasp-121_2016}, which were found using transit photometry from TRAPPIST \citep{Gillon2011} with both the Sloan-$z'$ filter ($\lambda_{eff}$ = 0.895 $\mu$m) and Johnson-$B$ filter ($\lambda_{eff}$ = 0.44 $\mu$m) and EulerCam \citep{Lendl2012} with the Gunn-$r'$ filter ($\lambda_{eff}$ = 0.664 $\mu$m) and Geneva-B filter ($\lambda_{eff}$ = 0.4265 $\mu$m), and the planetary radius reported was increased to account for deformation and asphericity of the planet. 
If we were to use the planet radius\footnote{The stellar radius has not varied greatly across studies and so we keep it fixed in our analysis.} found in that work for the post-processing it would increase the modeled $F_p/F_*$ values by 17$\%$.} 

\added{As for the wavelength-dependence of the planet's emitting area, future comparisons between models and data will want to calculate the predicted spectra more carefully than we have been able to do so here. In particular, \cite{Fortney2019} demonstrate how a correction for this geometry can be included in models that calculate radiative transfer in the radial direction (typically with a plane-parallel correction for slanted optical paths), such as the code used to calculate spectra from the SPARC/MITgcm models. Alternatively, code that calculates radiative transfer along line-of-sight paths (such as used to calculate spectra from the RM-GCM models) should intrinsically account for the wavelength-dependent emitting areas. However, either method still needs to have the planet radius normalized, as addressed above, and the most physically self-consistent approach \citep[advocated by][]{Fortney2019} would be to calculate transmission spectra from each model and then normalize the planet radius to match observed values. }

\added{Taking into account all of the complications discussed above about calculating simulated spectra from 3-D models, we still find that our models absorb and re-emit too much starlight, under-predicting reflection from the planet's atmosphere. This is complementary to the Energy Balance Model result by \cite{Splinter2025} that found an unexpectedly large Bond albedo for this planet ($A_B=0.28$) and one\footnote{In that work dayside emission spectra were best fit by models that considered either TiO or reflected light.} of the dayside spectrum solutions from \cite{Pelletier2025} that found a geometric albedo of $A_g = 0.16$. In comparison, the largest Bond albedo used in our set of 3-D models is $A_B = 0.05$ in the RM-GCM ones, with much smaller values for the SPARC/MITgcm ones ($<0.01$). Previous studies have found that the Bond albedo from Rayleigh scattering in clear H/He-dominated atmospheres will depend on temperature and metallicity \citep{parmentier_non-grey_2015,malsky_direct_2024}, but we should expect $A_B\lesssim 0.125$ for solar metallicity atmospheres and this decreases with increasing temperature and metallicity. The significantly higher albedo for WASP-121~b implied by these data continues an observational trend of unexpectedly high bond albedos across the hot Jupiter population compared to both geometric albedo and theoretical predictions \citep[with average observed values of $A_B \simeq 0.35$ and $A_g \simeq 0.1$;][]{schwartz_balancing_2015}.}

\added{From the NIRISS data alone, we can use the results from \cite{Splinter2025} and \cite{Pelletier2025} to determine a possible $A_B/A_g=1.7$ for WASP-121~b, keeping in mind that \cite{Pelletier2025} also found a solution with zero reflection. $A_B/A_g\approx1.3$ for Rayleigh scattering in a clear atmosphere \citep{Heng2021}, while a value of $A_B/A_g=1.7$ would imply the presence of dayside clouds, which is not predicted by our 3-D models. Cloud modeling is notoriously difficult, coupling physics across scales from microphysical to global, so perhaps these 3-D models are missing some microphysical mechanisms that would allow clouds to persist longer across the planet's dayside \citep{powell_two-dimensional_2024} and provide some extra reflectance. This is complimentary from the finding in \cite{Heng2026} that the large ratio between bond and geometric albedo can be explained by the presence of clouds at the terminator of hot Jupiters. However, increased presence of clouds in the western regions of the dayside means that the phase curve offset in reflected-light wavelengths should be in the opposite direction as observed in the NIRISS data (as discussed above in Section \ref{sec:waveoffsets}). In summary, we confirm a mismatch between the expected and observed (or implied) albedo for WASP-121~b and resolving this will require additional modeling and/or observational investigation.}

\subsubsection{Limitations of Picket-Fence Radiative Transfer in the UHJ Regime \label{sec:picket-fence}}

The picket fence radiative transfer used in the RM-GCM models (PfBase, etc.) results in dayside thermal structures that are too hot and too isothermal when compared to those found in correlated-k models and inferred from observations, as seen by the overly bright and relatively featureless dayside spectra (Figure \ref{fig:allday}). While picket fence radiative transfer has been shown to give comparable results to the correlated-k scheme in \cite{lee_simulating_2021}, these tests were for the cooler hot Jupiter HD~209458b. The similar amplitudes of the spectral features between the nightside spectra predicted by the correlated-k CkBase model and the picket fence PfMag model (Figure \ref{fig:allnight}) is another demonstration that the picket fence scheme can reproduce realistic thermal profiles in cooler regimes. However, for ultra-hot Jupiter planets, the picket fence radiative transfer more poorly captures the irradiated thermal structure, struggling to produce temperature inversions that are as large in temperature change and cover as extensive of a pressure range as models with correlated-k radiative transfer. The UHJ picket fence models often have thermal inversions deep into the atmosphere ($\sim$0.1-1 bar) and then are isothermal above this \citep{kennedy_radiatively_2025}, whereas in the correlated-k models the temperature inversions extend to very high in the atmosphere ($\sim 10^{-4}$ bar). We expect the reason for this difficulty in the picket fence method is that, while the absorption of starlight is functionally correct for picket fence when compared to correlated-k, the thermal channels within picket fence are not flexible enough at only two channels to properly capture the shape of UHJ radiative equilibrium profiles. 

While this means that the Pf models are not accurately capturing the inverted structure of the dayside, this does not negate what we can learn from these models about the impact of clouds and (magnetic) drag. Even with the lower dayside temperatures indicated by the data, we still expect the clouds on this planet to be confined to the nightside, where the picket fence treatment is more accurate. Dayside temperature inversions are not required for magnetic drag to occur, only temperatures sufficiently high enough for thermal ionization \citep{perna_magnetic_2010}. While these models also predict too hot of a dayside, the data still imply a hot enough dayside for there to be significant thermal ionization and, since the strength of the planet's global magnetic field is unconstrained, we cannot know how strong the drag timescales should be. In this way, PfMag models, in comparison to Pf models without drag, contribute to our interpretation of the planet has having strong drag operating in its atmosphere.

\section{Conclusion} \label{sec:conc}

In this work we have used a set of nine models produced from three-dimensional General Circulation Models (GCMs) to interpret the JWST/NIRISS spectroscopic phase curve of the ultra-hot Jupiter WASP-121~b \citep{Splinter2025}. With these models we have explored how this planet's atmosphere is shaped by different physical effects: the atmospheric opacities that determine the absorption of starlight, 
drag on the atmospheric winds, clouds, and the dissociation and recombination of H$_2$. With the precision and broad wavelength coverage of JWST/NIRISS, we are able to confirm results from previous observations of this planet and provide additional insight into its atmospheric state. Our key findings are:
\begin{itemize}
   \item \added{The measured emission from WASP-121~b is lower than predicted by most of the 3-D models, as seen in the white-light phase curve and dayside spectrum, although this is sensitive to the choice of planetary radius. This mismatch in emission highlights the need for careful consideration of radius when producing planet-to-star flux ratios and is complementary to the empirical findings from \cite{Splinter2025} and \cite{Pelletier2025} that this planet has a significant albedo (respectively: $A_B=0.277\pm0.016$ and $A_g=0.16\pm0.02$), although these values exceed theoretical expectations. }
   \item We confirm the findings from previous phase curve measurements of this planet that suggested a source of strong drag in its atmosphere. There are multiple pieces of evidence in the full NIRISS dataset that indicate this result: the close alignment between the brightest region of the planet and its substellar point, with this trend persisting as comparable levels of emission from symmetric phases (even into nightside phases), and the lack of any strong spectral variation in the phase curve offsets.
   \item Although the dimness of the nightside spectrum requires a strong source of drag and cannot be explained by the presence of clouds alone in the models considered, the lack of strong absorption features does point toward the likely presence of nightside clouds. It might be possible to explain the featureless spectrum as a balance in the hemispherically integrated flux between the cool anti-stellar region and the hot, inverted near-terminator region, but that scenario more poorly matches the previously measured JWST/NIRSpec nightside fluxes.
\end{itemize}

These results also identify a feature of the planet that remains mysterious and motivates future work, both observational and theoretical. At wavelengths shorter than $\sim1.4\mu$m the NIRISS data show an increase in the eastward phase curve offset with decreasing wavelength, which is not well matched by any 3-D models. We cannot feasibly explain this observed behavior with reflected light from clouds nor changes in the atmospheric opacity over these wavelengths. There may be some tension between the offsets measured in the NIRISS and TESS phase curves of this planet, but the uncertainties in the NIRISS data increase strongly at the shortest wavelengths.

This work highlights the complex nature of hot Jupiter atmospheres and how the precision of JWST is pushing models to a new ground, one where more physical phenomena must be considered and with higher complexity if we hope to explain, in detail, what we are observing. Despite this, GCMs are a uniquely useful tool for piecing together information available from different sides of these intrinsically 3-D objects into a cohesive picture of the planet's atmosphere state.

\begin{acknowledgments}
We thank the referee, Thaddeus Komacek, for his careful reading and constructive review, which resulted in significant improvements to the manuscript, including the identification of an issue the planet radius used in our modeled spectra.
R.C.F. and E.R. were partially supported on this work by Grant \#2019-1403 from the Heising-Simons Foundation. D.J.\ is supported by NRC Canada and by an NSERC Discovery Grant. S.P.\ acknowledges support from the Swiss National Science Foundation under grant 51NF40\_205606 within the framework of the National Centre of Competence in Research PlanetS. A portion of this research was carried out at the Jet Propulsion Laboratory, California Institute of Technology, under a contract with the National Aeronautics and Space Administration (80NM0018D0004). N.B.C. acknowledges support from an NSERC Discovery Grant, a Tier 2 Canada Research Chair, and an Arthur B.\ McDonald Fellowship and thanks the Trottier Space Institute and l'Institut de recherche sur les exoplan\'etes for their financial support and dynamic intellectual environment. R.A. acknowledges the Swiss National Science Foundation (SNSF) support under the Post-Doc Mobility grant P500PT\_222212 and the support of the Institut Trottier de Recherche sur les Exoplan\'etes (IREx). RJM is supported by NASA through the NASA Hubble Fellowship grant HST-HF2-51513.001, awarded by the Space Telescope Science Institute, which is operated by the Association of Universities for Research in Astronomy, Inc., for NASA, under contract NAS 5-26555. JDT acknowledges funding support by the TESS Guest Investigator Program G06165. L.D. is a Banting and Trottier Postdoctoral Fellow and acknowledges support from the Natural Sciences and Engineering Research Council (NSERC) and the Trottier Family Foundation.
This project was undertaken with the financial support of the Canadian Space Agency.

\end{acknowledgments}

\vspace{5mm}

\software{
    \texttt{astropy} \citep{astropy_collaboration_astropy_2013, astropy_collaboration_astropy_2018},
    \texttt{matplotlib} \citep{hunter_matplotlib_2007}, 
    \texttt{numpy} \citep{numpy}, 
    \texttt{pandas} \citep{pandas}, 
    \texttt{scipy} \citep{virtanen_scipy_2020}
}

\bibliography{references}{}
\bibliographystyle{aasjournal}


\section{Appendix}
\subsection{Additional Spectra Plots} \label{sec:appendplot}
\begin{figure*}[h]
    \centering
    \includegraphics[width=1\linewidth]{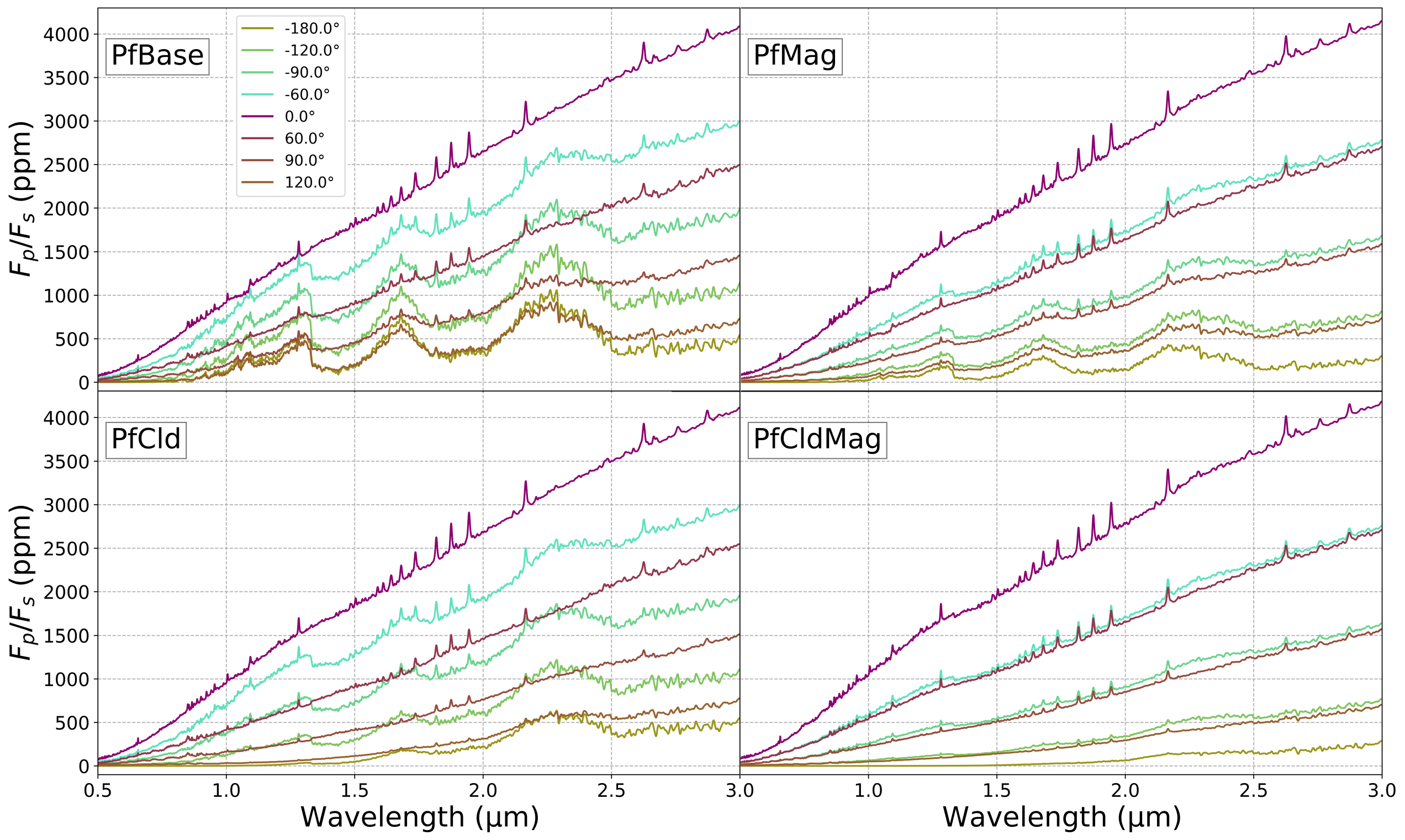}
    \caption{Predicted emission spectra of WASP-121~b, as would be observed at eight orbital phases including the dayside ($0\degree$) and nightside ($180\degree$) of the planet, from the four different models that used the RM-GCM, as labeled (see Table \ref{tab:gcmpres}). All models use picket fence radiative transfer. The PfMag and PfCldMag models apply a form of kinematic MHD drag. The PfCld and PfCldMag models include condensate clouds that actively form and dissipate as the model runs. None of these models include the effect of H$_2$ dissociation and recombination.}
    \label{fig:allrm}
\end{figure*}

\begin{figure*}
    \centering
    \includegraphics[width=1\linewidth]{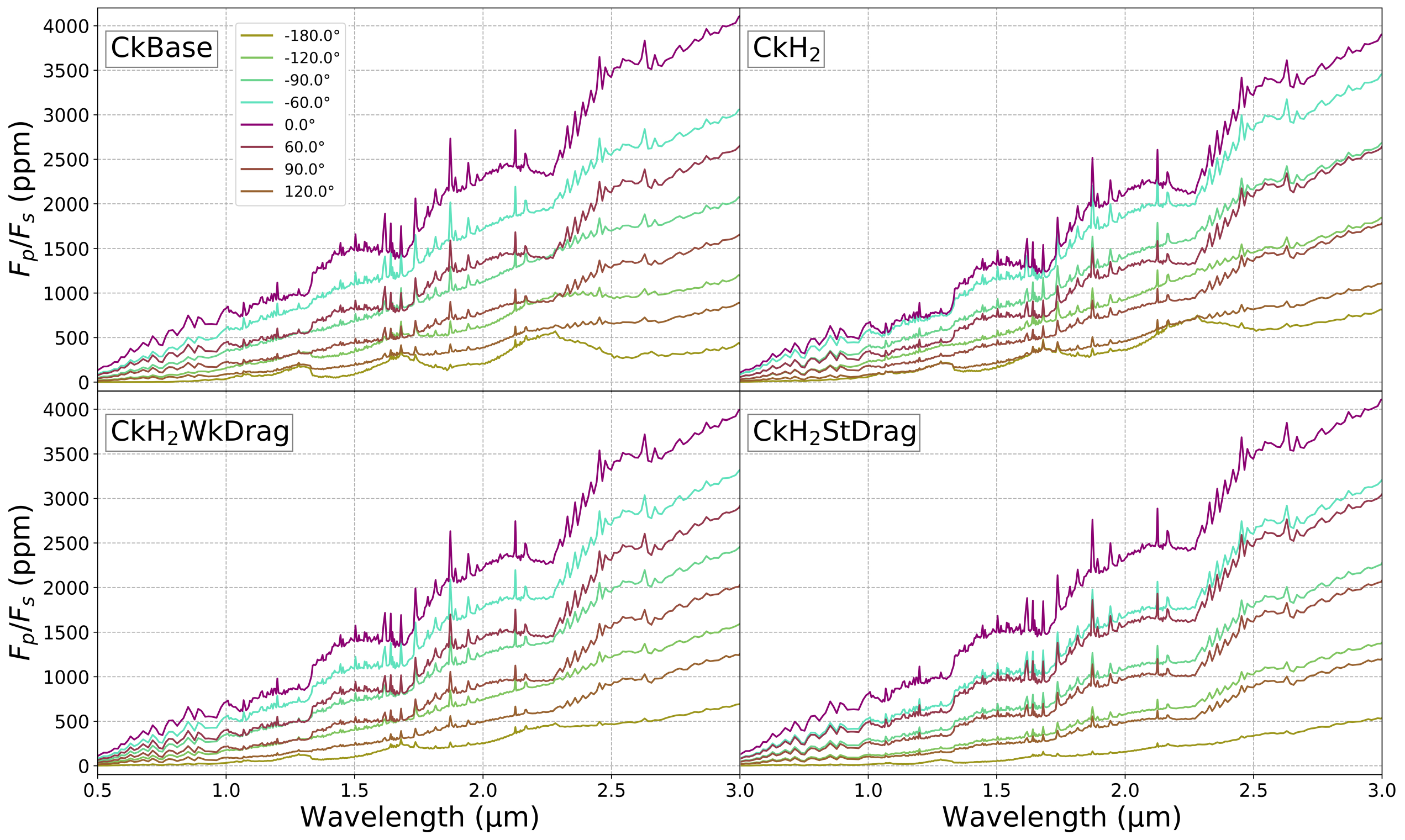}
    \caption{Predicted emission spectra of WASP-121~b, as would be observed at eight orbital phases, including the dayside ($0\degree$) and nightside ($180\degree$) of the planet, from the four different models that used the SPARC/MITgcm, as labeled (see Table \ref{tab:gcmpres}). All models use correlated-k radiative transfer. 
    The CkH$_2$ models include the effect of H$_2$ dissociation and recombination, while the WkDrag and StDrag versions also apply a drag with an uniform timescale (of $10^6$ or $10^4$ s, respectively). None of these models include clouds.}
    \label{fig:allsparc}
\end{figure*}

\begin{figure*}
    \centering
    \includegraphics[width=0.55\linewidth]{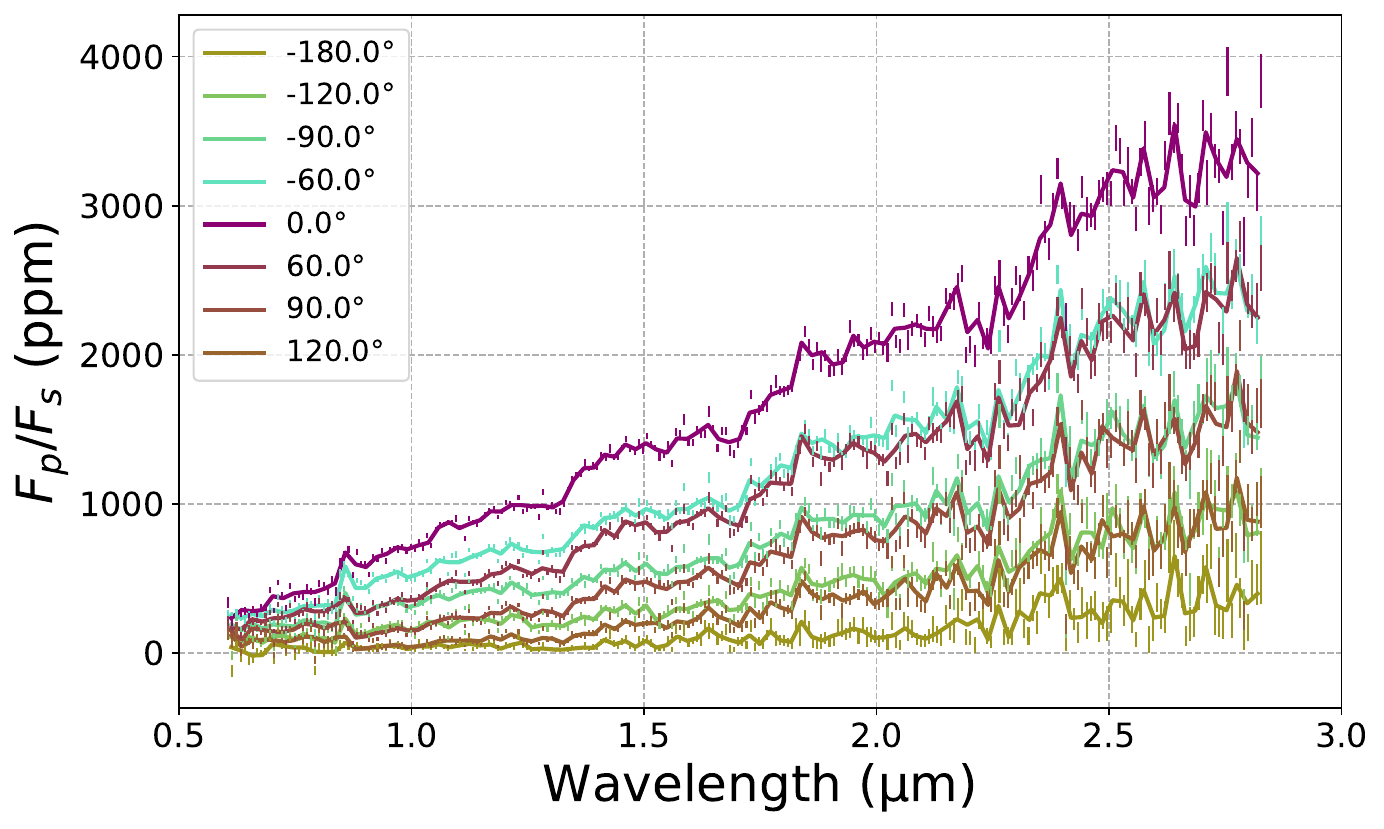}
    \caption{Emission spectra of WASP-121~b from NIRISS/SOSS at eight orbital phases, including the dayside ($0\degree$) and nightside ($180\degree$) of the planet, from the analysis performed in \cite{Splinter2025}. The spectra were initially produced binned down at 5 pixels; the lines plotted here are for the data binned down to 100 points and the error bars are for the data binned down to 250 points, from an initial 522.}
    \label{fig:allniriss}
\end{figure*}

Here we show the $F_p/F_*$ spectra calculated from each 3-D model as well as the NIRISS observations, across the NIRISS wavelength range and at eight evenly-sampled phases throughout the planet's orbit. Figure \ref{fig:allrm} shows the spectra from the models using the RM-GCM, Figure \ref{fig:allsparc} shows the spectra from those using the SPARC/MITgcm, and Figure \ref{fig:allniriss} shows the spectra from the exoTEDRF reduction of the NIRISS observation.

\subsection{Gas Opacities in RM-GCM Post-Processing} \label{gasopac}
This section contains a brief description on how gaseous opacities are accounted for in the most recent version of the RM-GCM post-processing pipeline. As stated in the Methods Section (\ref{sec:meth}), mixing ratios are calculated using the \texttt{FastChem} equilibrium chemistry code \citep{stock_fastchem_2018,stock_fastchem_2022}. We use a temperature-pressure grid that spans 200 to 5000 K and 1 $\times$ 10$^{-7}$ to 1000 bars. The wavelength-dependent gaseous opacities were calculated using the \texttt{HELIOS-K} \citep{grimm2015helios, grimm2021helios} opacity calculator. These opacities are then sampled using the \texttt{HELIOS} \citep{malik2017helios} k-table functionality. The post-processing code includes the following possible opacity sources (though these were not all used in this work): $\rm C$, $\rm CO$, $\rm AlH$, $\rm AlO$, $\rm C_2H_2$, $\rm Ca$, $\rm Ca+$, $\rm CaH$, $\rm CaO$, $\rm CH4$, $\rm CO2$, $\rm CrH$, $\rm FeH$, $\rm Fe$, $\rm H_2O$, $\rm H_2S$, $\rm HCN$, $\rm HF$, $\rm K$, $\rm Mg$, $\rm MgH$, $\rm Na$, $\rm NaH$, $\rm NH_3$, $\rm NO$, $\rm O_2$, $\rm O_3$, $\rm OH$, $\rm PH_3$, $\rm SH$, $\rm SiH$, $\rm SiH_4$, $\rm SiO$, $\rm SO_2$, $\rm TiO$, $\rm VO$ \citep{kurucz1995, dulick2003line, harris2006improved, yurchenko2014exomol, barber2014exomol, al2015marvel, li2015rovibrational, azzam2016exomol, mckemmish2016exomol, yurchenko2017hybrid, chubb2018marvel, polyansky2018exomol, mckemmish2019exomol, coles2019exomol, yurchenko2020exomol, chubb2020exomol, bernath2020mollist, somogyi2021calculation}, with corresponding abundances generated for the above species. The collisionally-induced opacity sources include the following species: $\rm H_2-H_2$, $\rm H_2-He$, $\rm H_2-H$, $\rm H_2-CH_4$, $\rm CH_4-Ar$, $\rm CH_4-CH_4$, $\rm CO_2-CO_2$, $\rm He-H$, $\rm N_2-CH_4$, $\rm N_2-H_2$, $\rm N_2-N_2$, $\rm O_2-CO_2$, $\rm O_2-N_2$, $\rm O_2-O_2$, $\rm H-e$ \added{(which accounts for H$^-$ opacities)}. Rayleigh scattering is also included, as previously described in \cite{zhang_constraining_2017}, and does not account for adding scattered light back in as emission. 

\subsection{Additional Model Pressure-Temperature Profiles} \label{sec:appendTPs}
\begin{figure*}[h]
    \centering
    \includegraphics[width=.6\linewidth]{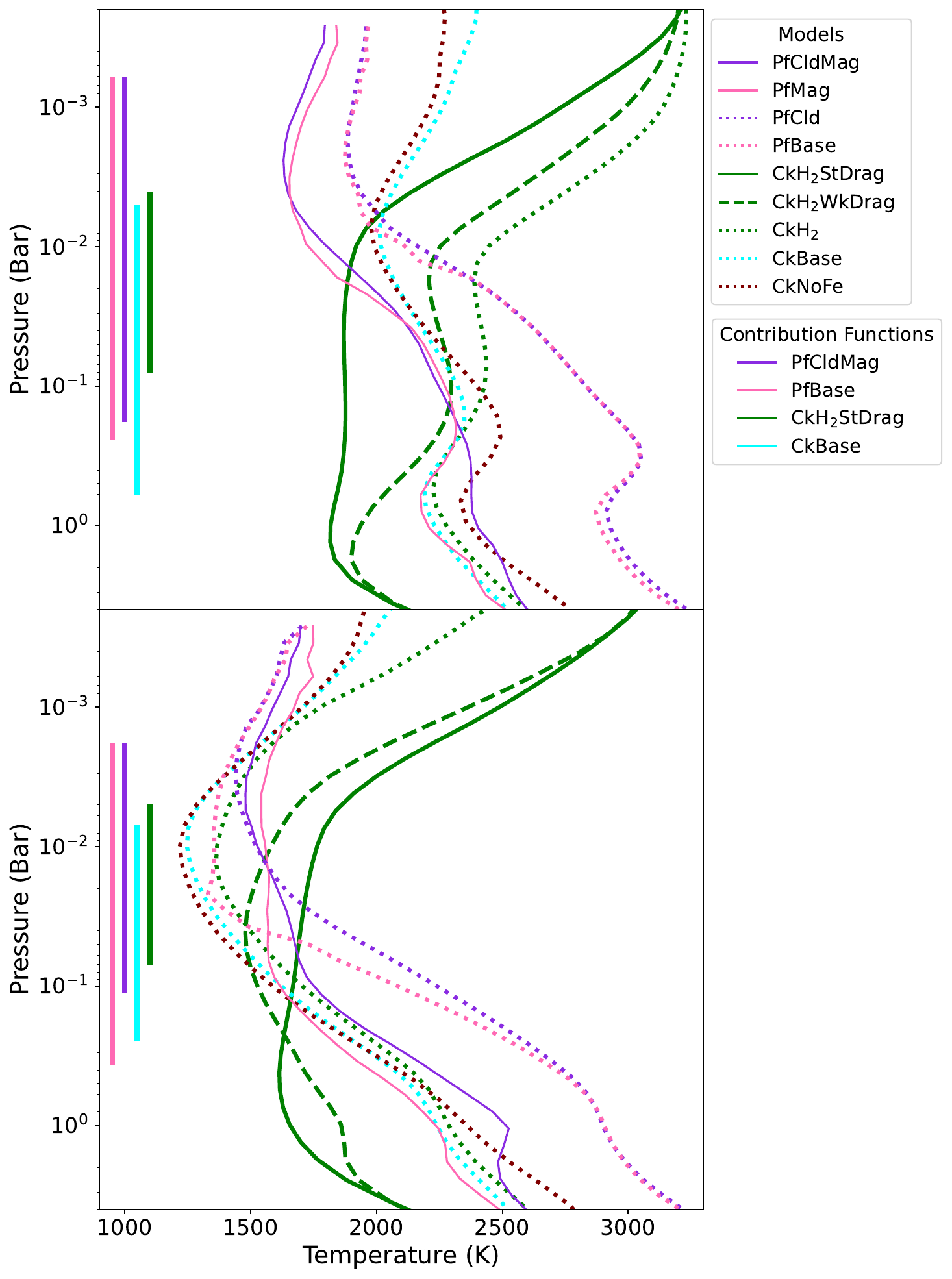}
    \caption{Spatially averaged pressure-temperature profiles of the 3-D models of WASP-121~b for phases $\pm90\degree$, regions where the day and nightside are in equal view but seen from opposite sides of the planet. The profiles are averaged over a 30$\degree$ region centered on a point of the specified longitude and the equatorial latitude. The vertical lines on the left-hand side of each plot indicate pressure ranges where observed emission originates from. The top panel shows profiles for phase -90$\degree$ and the bottom panel is phase +90$\degree$. Profiles from non-dragged models have significant differences, with notably hotter profiles to the east of the substellar point than to the west of it. Profiles from strongly dragged models have, in comparison, minimal change between the two regions. }
    \label{fig:symmetry}
\end{figure*}

Here we show pressure-temperature profiles from the nine models used in this work, for regions of the planet beyond those centered on the substellar and antistellar points (shown in Figure \ref{fig:day-nightPTs}). Figure \ref{fig:symmetry} shows averaged profiles for each model, for regions offset 90 degrees east and west of the substellar point, where the day and nightside are equally visible but from opposite sides. 

\end{document}